\documentclass[openacc]{rsproca_new}

\usepackage[numbers,sort&compress]{natbib} 
\usepackage{color}
\newcommand{\refAll}[1]{{\color{black}#1}}

\graphicspath{{img/}}
\newcommand\pd{\partial}
\newcommand\pdt{\partial_t}
\renewcommand{\vec}[1]{\boldsymbol{\mathrm{#1}}} 
\newcommand{\ten}[1]{\boldsymbol{\mathrm{{#1}}}}
\providecommand\bnabla{\boldsymbol{\nabla}}
\providecommand\bcdot{\boldsymbol{\cdot}}
\newcommand{\scr}[1]{\mathrm{#1}}
\newcommand{\ord}{\epsilon}
\newcommand{\ordest}[1]{\mathcal{O} \left( #1\right)}

\newcommand{\hn}{\hat{n}}
\newcommand{\htau}{\hat{\tau}}

\newcommand{\str}[3]{\varepsilon^{#1}_{#2} \left( #3 \right)}
\newcommand{\vare}[3][]{ #3^{#1(#2)} }

\begin{document}

\title{A computational continuum model of poroelastic beds}

\author{
U. L\={a}cis$^{1}$, G. A. Zampogna$^{2}$, S. Bagheri$^{1}$}

\address{$^{1}$Linn\'{e} Flow Centre, Department of Mechanics KTH, SE-100 44 Stockholm, Sweden\\
$^{2}$DICCA, Scuola Politecnica, Universit\`{a} di Genova, via Montallegro 1, 16145 Genova, Italy}

\subject{Biomechanics, computational mechanics, fluid mechanics}


\corres{Shervin Bagheri\\
\email{shervin@mech.kth.se}}

\begin{abstract} 
Despite the ubiquity of fluid flows interacting with porous and elastic
materials, we lack a validated non-empirical macroscale method for characterizing the flow over and through a poroelastic medium. We propose a  computational tool to describe such configurations by deriving and validating a continuum model for the poroelastic bed and its interface with the above free fluid.
We show that, using stress continuity condition and slip velocity condition
at the interface,
the effective model captures the effects of small changes in
the microstructure anisotropy correctly and predicts the overall
behaviour in a physically consistent and controllable manner.
Moreover, we show that the performance of the effective model is accurate by validating with fully microscopic resolved simulations.
The proposed
computational tool can be used in investigations in a wide  range of fields, including mechanical engineering, bio-engineering and geophysics.
\end{abstract}

\begin{fmtext}
\section{Introduction}

{Recent advances in surface micro- and nano-fabrication
techniques\cite{brodie1992physics,Fleck2495,vaezi2013review}
are providing new technological oppor-tunities of enormous potential. However, we lack high-fidelity models capturing how the underlying small-scale physicochemical processes interact with the large-scale flow and heat- and mass-transport phenomena. The reason is the vast range of scales in both time and space that need be resolved in order to capture the full physical picture,
which renders full-scale numerical investigations extremely
costly\cite{keyes2013multiphysics}.
The development of multi-scale models with reduced complexity is a necessary enabler in an increasing number of applications where the control of events at the small length scales determines the properties of the flow over much larger space and time scales. Examples include, but are not limited to, \unskip\parfillskip 0pt \par}
\end{fmtext}


\maketitle

%
\noindent the design of novel surfaces to control heat transfer, skin friction or pressure drag, acoustic noise and fluid mixing.

There is thus a clear need to develop techniques that combine computational fluid mechanics with mesoscopic or microscopic models of materials. In this work, we take a step in this direction by modelling the interaction of mesoscopic surface textures  with macroscopic flows. By ``mesoscopic'' we mean structures that can be described by continuum methods, but are still significantly smaller than the large-scale flow phenomena. When these constructions, which can take the form of high-aspect ratio structures (such as fibres) or granular-type structures (such as particles), are saturated with a fluid and can absorb the fluid stress and possibly deform, we have a poroelastic material.

The description of poroelastic medium as a continuum has an extensive and broad history, beginning with the empirical models of porous media by Darcy\cite{darcy1856fontaines} and followed by the formulation of total stress tensor of deformable porous media by Biot\cite{biot1941general,biot1956lowF}. Today, there exists a significant amount of
work\cite{lee1997thermal,lee1997re,murad2001micromechanical,iliev2008upscaling,
mei2010homogenization,simpson2010multiscale,simpson2010multiscaleP2,kalcioglu2012macro,
whitaker1986flow_Defrom,pride2003linear,Gajo3061,gajo2011finite,
eikerling2011poroelectroelastic,delavoipiere2016poroelastic,
Selvadurai2779,Selvadurai20150418,chan2012poroelastic,C6SM02111E,
kapellos2012multiscale,Lee2051,minale2014momentumP1,gopinath2011elastohydrodynamics} on poroelastic
media adopting top-down approaches based empirical macroscale stress/deformation tests, bottom-up approaches using homogenization and volume averaging techniques, physical approaches using analytical and mechanical models, \refAll{or other methods}.

In this work, we employ the method of homogenization via multi-scale expansion to model flows through and over poroelastic surfaces. Within the general multi-scale analysis platform (MAP) classification proposed by Scheibe et al.\cite{scheibe2015analysis}, the current method falls into the category of methods for which the different scales can be completely decoupled from each other (e.g. ``motif B'' in \cite{scheibe2015analysis}). In other words, our work is ``formal upscaling'', which means that after the macroscale governing equations have been derived, the parameters of the model are governed by fully decoupled microscale problems (from now on, we use the term ``microscale'' instead of
``mesoscale'' to conform with the terminology within two-scale expansion approach).
This method limits our investigations to set-ups for which 
the underlying microscale closure problems are linear.
For example, the fluid phase in the poroelastic material must be slow enough such
that inertial effects can be neglected or modelled through some kind of linearisation.
Essentially all previous upscaling works on deformable porous media has been in the same motif B category. Examples include the method of volume averaging by Whitaker\cite{whitaker1998method} and the method of homogenization, as employed, for example, by Mei \& Vernescu\cite{mei2010homogenization}.

There exists now a number of studies \citep[and references within]{cushman2002primer} on development and applications of the motif B multi-scale methods for poroelastic media.
It is, however, the authors' opinion that we still lack a computational framework that lays out the sequential steps needed to be taken in order to obtain the physical parameters describing the poroelastic medium not only in the interior of the material, but also at the interface with freely moving fluid. Ideally, such a framework, when provided a particular microscopic structure in terms of its geometry and properties (skeleton elasticity, connectivity, etc.), will provide the anisotropic macroscopic material properties (permeability, elasticity, etc.) of the effective continuum fields without any fitting parameters obtained from experiments.

To the best of the authors' knowledge, the theoretical frameworks of
Whitaker\cite {whitaker1986flow_Defrom} and Mei \& Vernescu\cite{mei2010homogenization}
have not been validated from microscopic point of view for the flow over
a surface, which is coated with porous and elastic material. 
Validations of these methods most often consider only macroscopic measures.
More specifically, while microscale problems in unit cells needed in the upscaling procedures have been presented and solved
previously\cite{whitaker1986flow_Defrom,lee1997thermalP2,mei2010homogenization}, there is no comparison between global macroscale simulations -- using properties obtained from those microscale solutions -- and corresponding fully resolved simulations. In this work, we do not only compute effective material tensors by numerically solving microscale problems, but also compare the obtained effective continuum description with fully resolved direct numerical simulation (DNS) of the fluid flow inside and above the poroelastic surface. In this way, we can assess the accuracy of the effective model quantitatively both in the interior and near the interface of the medium with a free-flowing fluid.

A particularly important contribution
is the insight it provides on how the velocity and the stress interface conditions between the poroelastic medium and the free fluid region model the microscale effects
in averaged manner. Gopinath \& Mahadevan\cite{gopinath2011elastohydrodynamics} as well as Minale\cite{minale2014momentum} point out that the effective interface condition for a poroelastic region may in fact be more straightforward to match, compared to the rigid porous material, since there is a natural way to balance the fluid stress from the free fluid with the solid stress of the surface material. However, there exists no validation -- where the interface boundary conditions for displacement, fluid velocity and pressure are coupled to the Stokes equations above --
for a non-trivial flow, where there is transport across the interface. Most works treating the boundary conditions are empirical\citep{han2005transmission,le2006interfacial,rosti2015direct,bottaro2016rigfibre}. Those contributions which have recently treated the interface problems from first-principles have focused on rigid porous media and one-dimensional problems, such as the laminar channel flow \cite{mikelic2000interface,auriault2010beavers} only (for which there is no transfer of mass or momentum between the material and the free fluid), or infiltration flow only\cite{carraro2015effective}.

In summary, the objectives of the current work are to (i) present a framework, derived using multi-scale expansion, suitable to model flow through and over poroelastic materials, (ii) validate the framework with respect to the
fully resolved direct numerical simulations,
and (iii) gain insight on stress transfer near the interface and evaluate the accuracy of interface boundary conditions. Our work is partially the numerical counterpart of the analytical and asymptotic study by Gopinath \& Mahadevan\cite{gopinath2011elastohydrodynamics}; we aim at computing (instead of physically modelling) physical material properties of anisotropic poroelastic materials as well as the interface with the free flow. While 
Gopinath \& Mahadevan\cite{gopinath2011elastohydrodynamics} consider particularly biologically relevant microstructures of ordered or disordered filaments, we consider connected materials consisting of linked spheres and ellipsoids.

This paper is organized as follows. In section~\ref{sec:homog-eq}, we introduce both the microscale and the macroscale/effective equations governing our problem, which consists of a poroelastic material at the interface with a moving fluid.
A method to compute the effective properties of poroelastic materials is presented
in section~\ref{sec:effective-prop}.
In the same section, we provide effective tensor results
for cubic-symmetric and monoclinic-symmetric poroelastic materials and analyse them.
A lid-driven cavity problem to investigate poroelastic material response to a steady
two-dimensional flow vortex is proposed and solved using the homogenized
equations in section~\ref{sec:results-cavity}. In the same section, we 
report results obtained from resolved direct numerical simulations
\refAll{and explain
the shear stress transfer between free fluid and poroelastic medium}.
In section~\ref{sec:discuss}, we discuss the limitations of the presented
theory.
Finally, in section~\ref{sec:concl} we conclude this work and
outline future directions.

\section{Micro- and macroscale equations describing a poroelastic bed} \label{sec:homog-eq}
\begin{figure}
  \begin{center}
  \includegraphics[width=0.69\linewidth]{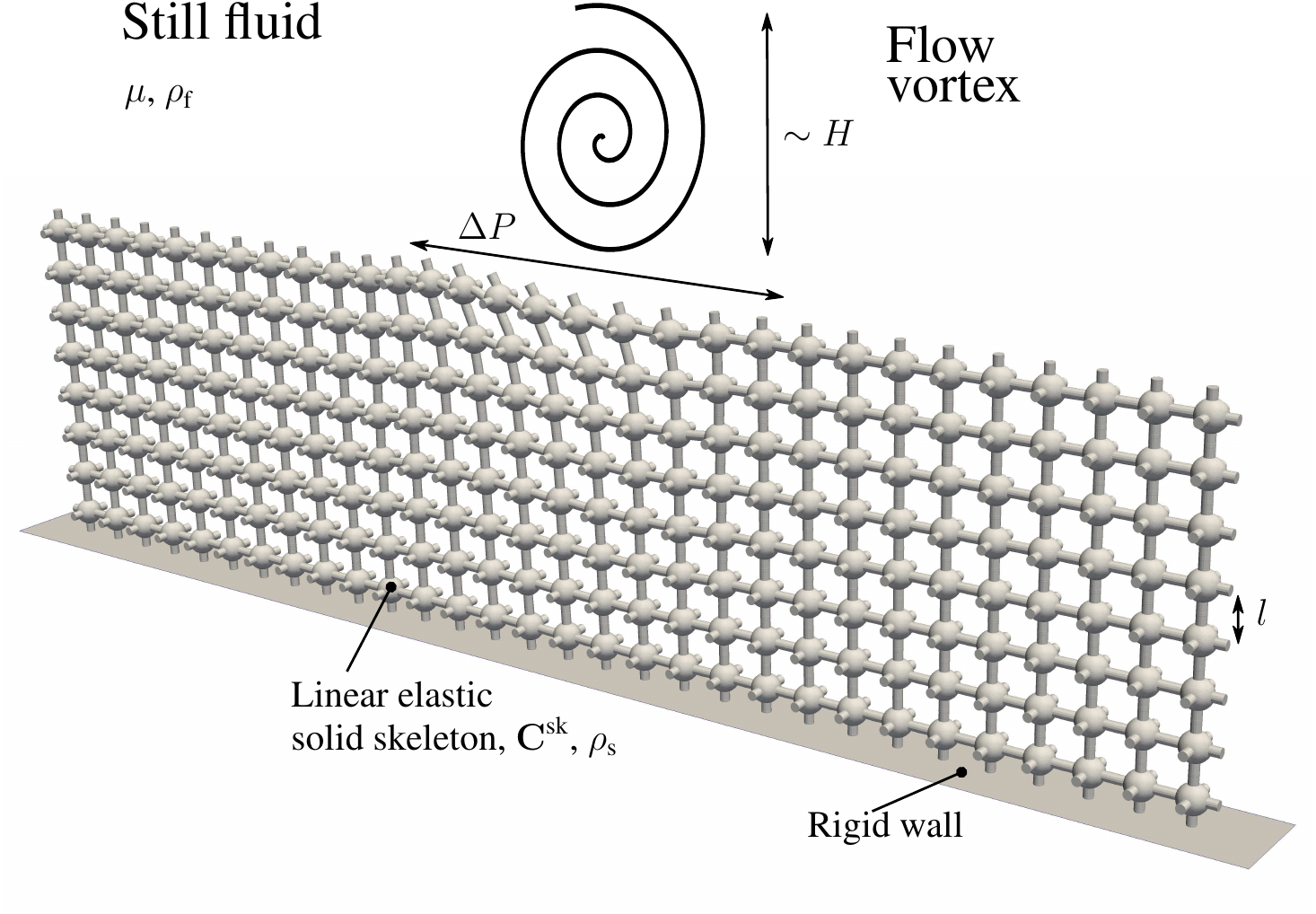}
  \vspace*{-20pt}
  \end{center}
  \caption{ 
Illustration of a free fluid vortex interaction with a poroelastic material.
The material is composed of multiple instances of unit-cell skeleton
geometry -- sphere with circular cylinder connections in all directions. 
We represent a slice one pore-structure thick. The unit-cell is a cube with
side length $l$. The solid skeleton is characterized by linear elasticity
tensor $\ten{C}^\scr{sk}$ \refAll{and density
$\rho_\scr{s}$}. The flow vortex has a length scale $H$ and causes a
characteristic pressure difference $\Delta P$. The solid skeleton is deformed
under the influence of the free fluid vortex.
\label{fig:pemedia-sketch}}
\end{figure}
For a dense poroelastic medium exposed to a free flow (Fig.~\ref{fig:pemedia-sketch}), one may define at least two length scales; a microscopic (pore) scale $l$, which characterizes the size of voids in the material and a macroscopic (global) scale $H$, which characterizes the size of the large-scale processes in the medium or nearby.
The large scale flow is characterized also by the created pressure difference
$\Delta P$.
In this section, we present the microscale equations that resolve every scale in the full physical domain, and a set of macroscale  equations that model the effective average behavior of the poroelastic bed under a given set of assumptions.  Although the effective field equations for describing a fluid-saturated poroelastic material are known \cite{biot1941general, gopinath2011elastohydrodynamics}, we re-derive them from first principles in the supplementary appendices. The main reason is to uncover the detailed  microscale problems in unit cells that are required for determining the physical coefficients appearing in the macroscale systems. 
%

\subsection{Microscale governing equations}

The microscale equations resolving the fluid-structure physics at scale $\sim l$ are the conservation of mass and momentum. For a Newtonian fluid with density $\rho_\scr{f}$ and viscosity $\mu$, the momentum in the free fluid region and within the porous skeleton is governed by the incompressible Navier-Stokes equations
\begin{align}
\rho_\scr{f} \left[ \pdt \vec{u} + \left( \vec{u} \cdot \nabla \right) \vec{u} \right]
& = -  \bnabla p + \mu \Delta \vec{u}, &  &
\label{eq:NStokes-ms-dimensional-1} \\
\bnabla \bcdot \vec{u} & = 0, \label{eq:NStokes-ms-dimensional-2}
\end{align}
where $\vec{u}$ and $p$ are velocity and pressure fields, respectively.

The  material is defined by the solid skeleton density $\rho_\scr{s}$ and linear elasticity tensor  $\ten{C}^{\scr{sk}}$. Assuming an isotropic material the elasticity tensor is defined by Young's modulus $E$ and Poisson's ratio $\nu$.  The  solid skeleton momentum is governed by a balance between solid inertia and stress, obtained using the linear stress-strain relationship
\begin{align}
\rho_\scr{s} \pdt^2 \vec{v} & = \bnabla \cdot \left\{ \ten{C}^{\scr{sk}} :
\frac{1}{2} \left[ \bnabla \vec{v} +
\left( \bnabla \vec{v} \right)^T \right]  \right\}, \label{eq:solid-ms-dimensional}
\end{align}
where $\vec{v}$ is the displacement field of the solid skeleton. In order to couple the fluid and structure problems, the no-slip condition and continuity of stress is prescribed at the boundary between solid skeleton and surrounding fluid, i.e.,
\begin{equation}
\vec{u} = \pdt \vec{v} \qquad \mbox{and} \qquad \left\{ - p \ten{\delta} + \mu\left[
\bnabla \vec{u} +
\left( \bnabla \vec{u} \right)^T \right] \right\} \cdot \hn = \left\{ \ten{C}^{\scr{sk}} :
\frac{1}{2} \left[ \bnabla \vec{v} +
\left( \bnabla \vec{v} \right)^T \right]  \right\} \cdot \hn. \label{eq:solid-fluid-ms-dimensional-bcs}
\end{equation}
Here, $\ten{\delta}$ is the second-rank identity tensor and $\hn$ is unit-normal vector at the boundary. Solving the governing equations everywhere at the pore scale is computationally very expensive due to a globally large domain and due to requirement of fine resolution near the pores. This motivates the development of an alternative continuum description, where the pore fluid and solid are considered as one composite, which given appropriate equations and constitutive relations presents the average behavior of the actual poroelastic bed.

\subsection{Effective field equations}

\begin{figure}
  \begin{center}
  \includegraphics[width=0.69\linewidth]{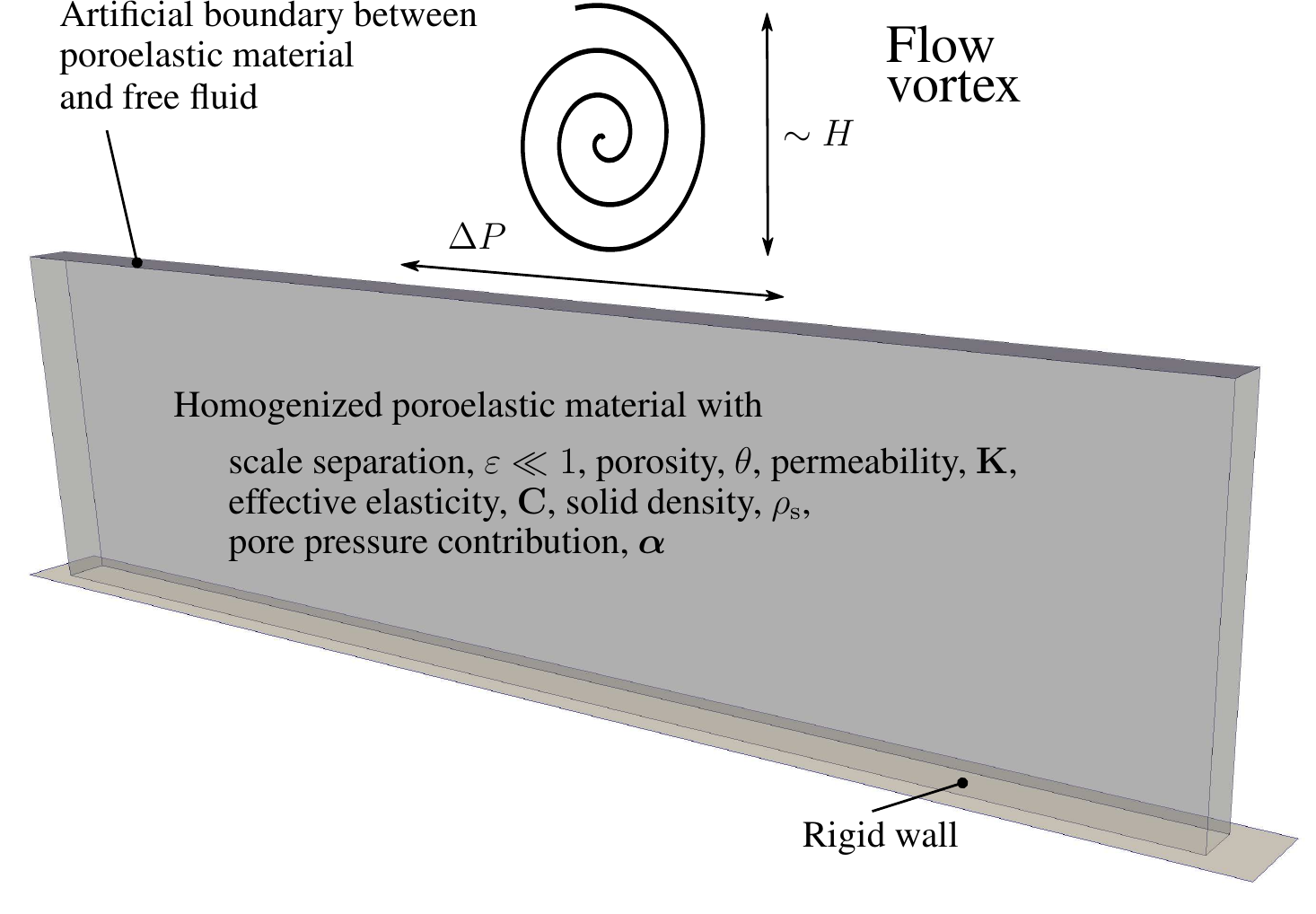}
  \vspace*{-20pt}
  \end{center}
  \caption{
Illustration of a free fluid vortex interaction with a poroelastic material.
The material is described in a homogenized two-domain
setting and is characterised by the scale
separation parameter $\ord$ and porosity $\theta$, as well as effective
permeability $\ten{K}$, effective elasticity $\ten{C}$,
\refAll{density $\rho_\scr{s}$} and pore pressure
contribution $\ten{\alpha}$. 
The homogenized domain is not
visibly deformed
under the influence of free fluid vortex. There is an artificial boundary
introduced between the homogenized poroelastic material and free fluid.
\label{fig:pemedia-sketch-mod}}
\end{figure}

We divide the physical domain into two parts; one containing only the free fluid, and the other containing the fluid and solid skeleton (Fig.~\ref{fig:pemedia-sketch-mod}). The free-fluid region is governed by the Navier-Stokes equations (\ref{eq:NStokes-ms-dimensional-1}--\ref{eq:NStokes-ms-dimensional-2}).

The continuum description of the composite (solid and fluid) is based on a
separation between the pore-scale and the system scale. Mathematically this can be formulated by a scale separation parameter $\ord = l/H \ll 1$.
%
%
%
The effective displacement field $\vec{v}$ of the homogenized poroelastic material (see Fig.~\ref{fig:pemedia-sketch-mod}) is governed by,
\begin{align}
\left( 1 - \theta \right) \rho_\scr{s}\, \pdt^2 \vec{v} & = \bnabla \cdot
\left[ \ten{C} : \frac{1}{2} \left( \bnabla \vec{v} +
\left( \bnabla \vec{v} \right)^T \right) - \ten{\alpha} p \right].
\label{eq:dim-pore-momentum}
\end{align}
Here, $\theta = V_\scr{f}/V$ is the porosity (which in general is a function of space, but in this work is a constant), where $V_\scr{f}$ is the fluid phase volume and $V$ is the total volume of the composite in one unit cell (defined later).
Moreover, $\ten{C}$ is the fourth-rank effective elasticity tensor of the material and $\ten{\alpha}$ is the coefficient for the contribution of the pore pressure ($p$) in the total stress. Whereas the microscale elasticity $\ten{C}^{\textrm{sk}}$ only depends on $E$ and $\nu$, the effective tensor $\ten{C}$ in general depends on the porosity, pore geometry
(as modelled in the current work) and on the particular type of boundary condition imposed near the interface with the free fluid, or the impermeable walls\cite{gopinath2011elastohydrodynamics}. The tensor $\ten{\alpha}$ does not have a correspondence in the microscale and is an effect
of solid skeleton deformation due to seepage flow through the pores. 
We characterize in detail both $\ten{C}$ and  $\ten{\alpha}$ for two different poroelastic media in section~\ref{sec:effective-prop}.
%

Moreover, 
for sufficiently dense poroelastic material the fluid flow between the pores is slow, such that inertial effects are negligible.
Therefore the pore pressure is the dominant contribution in the fluid flow and
the leading-order equation is the (relative) Darcy's law,
\begin{equation}
\vec{u} - \theta \pdt \vec{v} = - \frac{\ten{K}}{\mu}
\cdot \bnabla p. \label{eq:dim-rel-Darcy-law}
\end{equation}
This expression relates the gradient of the pore pressure, the solid velocity $\pdt \vec{v}$ and the flow field $\vec{u}$ in the poroelastic medium to each other. Here  $\ten{K}$ is the interior permeability tensor. The term $ \theta \pdt \vec{v}$ arises due to viscous friction between the solid structure and the pore fluid; if there is a motion of the solid skeleton, the surrounding pore fluid is dragged along through boundary condition (\ref{eq:solid-fluid-ms-dimensional-bcs}).

Finally, the conservation of mass requires that
\begin{align}
\bnabla \cdot \vec{u} & = \ten{D} : \frac{1}{2} \pdt \left( \bnabla \vec{v} + \left( \bnabla \vec{v} \right)^T \right) - \mathcal{E} \pdt p, \label{eq:dim-mass-coserv-vs-displace}
\end{align}
where $\ten{D}=\theta \ten{\delta} - \ten{\alpha}$ is a dimensionless second-rank 
tensor. It determines how the strain of the displacement (caused either by the flow
through the pores or by a boundary condition) modifies the solid structure volume within
one pore, thus squeezing the pore fluid in or out of the pore.
%
The scalar $\mathcal{E}$ characterizes the change of solid structure volume
within one pore
with respect to time varying pressure, which, similarly to solid strain, can cause
change in pore fluid content and consequently introduce apparent compressibility
of the flow field.

The system of equations (\ref{eq:dim-pore-momentum}-\ref{eq:dim-mass-coserv-vs-displace}) determines the seven unknowns ($\vec{u},\vec{v}$ and $p$), but by  combining (\ref{eq:dim-rel-Darcy-law}) with (\ref{eq:dim-mass-coserv-vs-displace}), the system can be reduced to four unknowns ($\vec{v}$ and $p$). The fluid velocity $\vec{u}$ can then be computed as a postprocessing step once pressure and displacement fields are known. Following this approach, the equation for the pore pressure is
\begin{align}
\mathcal{E} \pdt p - \bnabla \cdot \left(\frac{\ten{K}}{\mu} \cdot \bnabla p \right) & = -\ten{\alpha} : \frac{1}{2} \pdt \left( \bnabla \vec{v} + \left( \bnabla \vec{v} \right)^T \right). \label{eq:dim-pore-pressure}
\end{align}
%
%

Note that the effective system above for describing a poroelastic system can be derived using multi-scale expansion (see section~\ref{sec:effective-prop} and
supplementary appendices), motivated using mixture theory, or physically modelled.
The major challenge is to use this developed effective system with appropriate
boundary conditions at interfaces with solid walls, free fluids or other structures
to describe problems arising from various applications. 
The aim of this work is to provide a framework from which one can form a fully closed effective 
system to describe the response and interaction of a poroelastic material with a
surrounding free fluid. In the next section, we therefore provide the needed boundary conditions between a poroelastic material and a rigid wall, \refAll{and also} a free fluid.

\subsection{Effective interface conditions}


\subsubsection{Conditions for the poroelastic bed}
To solve the governing equations for the poroelastic material, one needs to impose
boundary conditions for both the pore-pressure (equation \ref{eq:dim-pore-pressure})
and the displacement field (equation \ref{eq:dim-pore-momentum}). 
On  rigid walls, we impose (similar to \cite{gopinath2011elastohydrodynamics}) zero displacement and  zero transpiration (normal fluid velocity), which through the relative Darcy's equation (\ref{eq:dim-rel-Darcy-law}) leads to
\begin{equation}
\left. \vec{v} \right|_\scr{rw} = 0 \qquad \mbox{and} \qquad
\left. \frac{\pd p}{\pd \hn} \right|_\scr{rw} = 0. \label{eq:homog-bc-walls}
\end{equation}
Here, ``$\scr{rw}$'' means ``rigid wall'' and $\hn$ is the unit-normal vector at the wall. Physically, the no-slip condition should be satisfied at the wall, but this is not compatible with the leading-order presentation of a poroelastic media based on Darcy's law. The Darcy's law only describes the direct proportionality between the pore-pressure gradient and the velocity, and does
not include any macroscopic diffusion effects.

At the artificial interface with the free fluid, shown in Fig.~\ref{fig:pemedia-sketch-mod},
we impose a pressure continuity condition
\begin{equation}
p^{-} = p. \label{eq:homog-p-bc-if}
\end{equation}
Here, the pore pressure below the interface is denoted by $p^-$ and pressure of the free fluid above the interface with $p$. Note that the choice of condition for pressure depends on the assumptions made about the flow
as well as the material geometry.
%
For example, L\={a}cis \& Bagheri\cite{lacis2016framework} have shown that, if the interface velocity
caused by the shear stress is of the same order as the velocity induced by the pore pressure gradient, the pressure continuity is the leading order boundary condition for any pore geometry.
On the other hand,
Mikeli\'c et al.\cite{marciniak2012effective,carraro2013pressure} have shown that, if velocity
contribution from the shear stress
at the interface is one order higher than contribution from the pressure gradient,
there is a pressure jump for anisotropic pore geometry. However, if the pore
geometry is isotropic/cubic-symmetric, the pressure continuity is still applicable.
%
Finally, for the material displacement, we impose continuity of stresses at the interface
\begin{equation}
\left[ \ten{C} : \frac{1}{2} \left( \bnabla \vec{v} +
\left( \bnabla \vec{v} \right)^T \right) - \ten{\alpha} p \right] \cdot \hn = 
\left[ - p \ten{\delta} + \mu \left( \bnabla \vec{u} + \left( \bnabla \vec{u} \right)^T
\right) \right] \cdot \hn. \label{eq:homog-v-bc-if}
\end{equation}
%
We thus assume, similarly to \cite{gopinath2011elastohydrodynamics},  that the total stress of the free fluid is transferred to the total effective stress of the interior poroelastic medium.
In general, the effective elasticity of the composite near the interface could however be different
from its value in the interior (it is argued in general to depend on
boudary conditions\cite{gopinath2011elastohydrodynamics}) and, to arrive to a more accurate boundary condition, one could construct an interface-cell with an elasticity problem, similarly as done for the velocity boundary condition\cite{lacis2016framework}. One objective of this paper is to understand how the fluid shear stress is transferred to the solid stress (first across the interface then inside the bed) and if the interface correction is necessary; in section~\ref{sec:results-cavity}, we will show that approximation of the interface effective stress with the interior parameters is able to capture the transfer of stress reasonably well.

\subsubsection{Conditions for the free fluid}
To solve for the Navier-Stokes equations in the free fluid domain, we need to
impose velocity -- in principle, one could also have stress condition, but it is already
used for the displacements of the poroelastic material -- boundary conditions at the interface.
%
In this work, we extend the velocity boundary condition derived by L\={a}cis \& Bagheri\cite{lacis2016framework} for a rigid porous bed to include poroelasticity\footnote{The discussion can be found in supplementary
Appendix~\ref{app:sec-app:deriv}, after equation (\ref{app:eq:lin-assumpt-fluid2}).}.
The condition for the tangential
interface velocity
is
\begin{equation}
\vec{u} \cdot \htau = \pdt \vec{v} \cdot \htau 
+ \left ( - \frac{\ten{K}^\scr{if}}{\mu} \cdot
\nabla p^{-} + \ten{L} : \left[ \nabla \vec{u}
+ \left( \nabla \vec{u} \right)^T \right] \right )
\cdot \htau, \label{eq:dim-vel-if-condition}
\end{equation}
%
where the unit vector $\htau$ denotes both tangential directions of the surface.
Note that the pressure gradient is the pore-pressure gradient from poroelastic material side of
the interface (hence the minus superscript), whereas the flow velocity field
$\vec{u}$
is on the free-fluid side.
The interface velocity has two distinct contributions; i) the no-slip contribution governed by 
the movement of the solid structure; ii) the slip contribution, which is caused by the porosity
of the solid structure and depends both on pore pressure gradient and free-fluid shear.
The slip contribution is characterized by the second-rank interface permeability
tensor $\ten{K}^\scr{if}$
and the third-rank slip length tensor $\ten{L}$.
For a dense material, the first slip term scales as $l^2$ and is significantly
smaller than the second term, which scales as $l$.
The motion of the poroelastic material or the no-slip contribution
depends not only on the
pore-length $l$, but also on the elasticity of the material and the flow regime.
%
The condition for the normal ``penetration'' interface velocity component is
set by mass conservation, i.e.
%
\begin{equation}
\vec{u} \cdot \hn = \pdt \vec{v} \cdot \hn - \left( \frac{\ten{K}}{\mu}
\cdot \nabla p^{-} \right)
\cdot \hn, \label{eq:dim-vel-if-condition-normal}
\end{equation}
where $\ten{K}$ is the interior permeability tensor defined in equation (\ref{eq:dim-rel-Darcy-law}) of the porous medium and $\nabla p^-$ denotes the pressure gradient when approaching the interface from the bed. Similarly as for the tangential component, the velocity has two parts -- 
the no-slip part and the ``slip'' part. The slip in the normal direction is essentially
the fluid mass transport in and out of the poroelastic material, which has to be equal to the
relative velocity from the interior. This same condition also arises from the interface
cell\cite{lacis2016framework}, because the tensor $\ten{L}$ components corresponding to
penetration interface velocity are zero.

In the next section, we will compute the physical parameters which characterize the interior poroelastic medium ($\ten{C}$, $\ten{\alpha}$, $\ten{K}$ and $\mathcal{E}$) as well as the parameters characterizing the interface with the free fluid ($\ten{K}^\scr{if}, \ten{L}$) for  two specific pore-scale geometries. In section \ref{sec:results-cavity},
\refAll{we will investigate in more details interface conditions
at boundary between poroelastic material and
free fluid by comparing the effective model
against the microscopically-resolved system using the lid-driven cavity problem.}

\section{Properties of poroelastic material exposed to free fluid} \label{sec:effective-prop}

We now turn to  determining the effective properties of a poroelastic material  exposed to the free fluid, by introducing and solving a set of particular microscale problems in an interior unit-cell and in an interface-cell. We focus on a periodic solid structure that can be obtained by duplicating a single pore structure in all  directions. The sketch in Fig.~\ref{fig:pemedia-design-def-cubicSym}$a$ shows how the material and its interface with free-fluid is divided into cubic interior cells and elongated rectangular interface-cells. The effective parameters are computed by solving two elasticity problems (to obtain $\ten{C}$, $\ten{\alpha}$ and $\mathcal{E}$) and one fluid problem (to obtain $\ten{K}$) in the interior cell and two fluid problems in the interface-cell (to obtain $\ten{K}^\scr{if}, \ten{L}$). These microscale problems are first illustrated for the interior domain using a weakly and a strongly anisotropic microstructure, before we move on to the microscale problems at the interface with the free fluid. 

\subsection{Interior of the poroelastic material} \label{sec:interior-prop}
\subsubsection{Microscale problems for elasticity parameters} \label{sec:cubic-sym-mat}
\begin{figure}
  \begin{center}
  \includegraphics[width=0.8\linewidth]{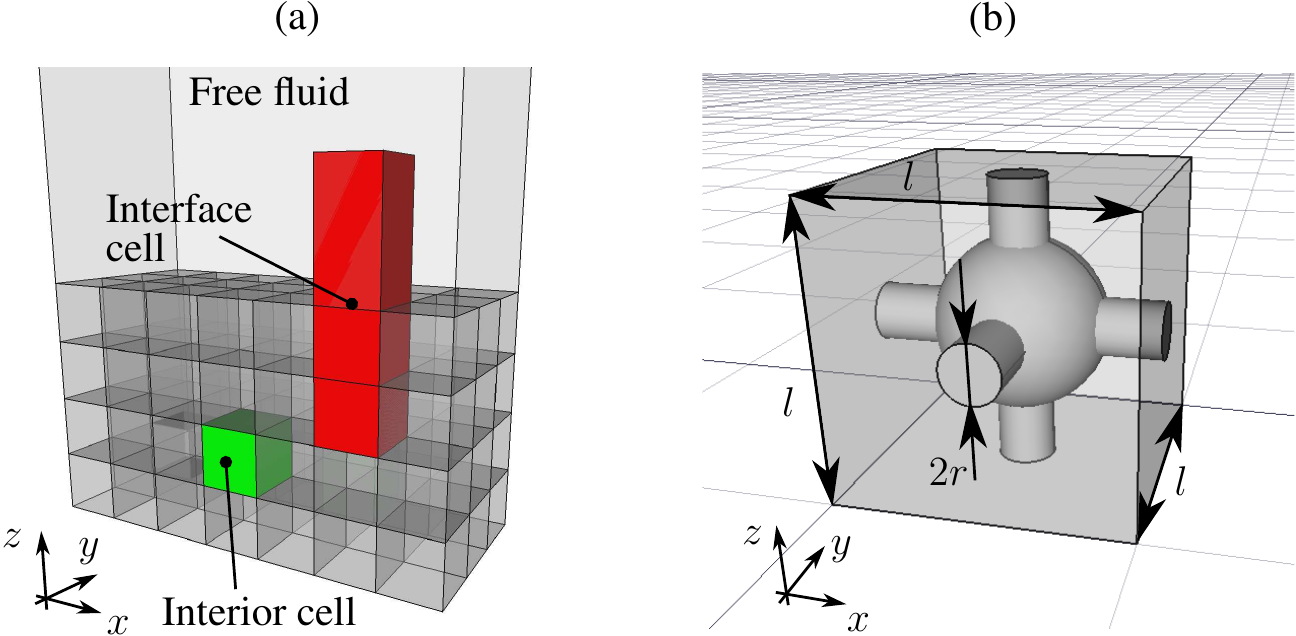} 
  \end{center}
  \caption{
In frame (a) we show a constructed poroelastic material consisting
of $84$ unit-cells with volume $l^3$, which can exhibit only cubic-symmetry
due to boundaries between
the unit-cells. \refAll{Above the material, there is a free fluid.
Interior unit-cell is depicted with green, interface-cell is depicted
with red.} In frame (b) we show a cubic-symmetric micro-structure placed in a
single interior unit-cell. The structure is built
using a sphere with radius $R = 0.3\,l$ and
cross-connected cylinders with radius $r = 0.1\,l$. 
\label{fig:pemedia-design-def-cubicSym}}
\end{figure}

The effective poroelastic bed obtained from the unit-cell approach can not have fully isotropic elastic properties due to the boundaries between cubic unit cells, as shown in Fig.~\ref{fig:pemedia-design-def-cubicSym}a. The resulting effective elasticity $\ten{C}$ will at most exhibit a cubic symmetry; a symmetry, which is defined by planes -- parallel to the unit-cell sides and diagonal
across the unit-cell in all directions --
going through the center of the cube. 

In order to preserve cubic symmetry, the unit-cell can be filled with any structure that itself is cubic symmetric, such as Wigner-Seitz grains \cite{lee1997thermalP2,mei2010homogenization} or cube with spherical holes \cite{lydzba2000study}. In this work, we use a sphere at the center of unit-cell with radius $R$, which is connected to neighbouring cells via circular cylinders of radius $r$, as shown in Fig.~\ref{fig:pemedia-design-def-cubicSym}b.  The solid skeleton structure is built from isotropic elastic
material with elasticity tensor
\begin{equation*}
C^{\textrm{sk}}_{ijkl} = E\left\{ \frac{\nu}{\left( 1 + \nu \right) \left( 1 - 2 \nu \right) } \delta_{ij} \delta_{kl} + \frac{1}{2 \left( 1 + \nu \right)} \left( \delta_{ik} \delta_{jl} + \delta_{il} \delta_{jk} \right) \right\}.
\end{equation*}
The effective elasticity tensor can be determined (see Appendix~\ref{app:sec-app:deriv} for derivation) using a third-rank displacement test tensor $\ten{\chi}$ as
\begin{equation}
\ten{C} = \left( 1 - \theta \right) \ten{C^\scr{sk}} + 
\ten{C^\scr{sk}} : \left\langle \frac{1}{2} \left[ \bnabla \ten{\chi}
+ \left(\bnabla \ten{\chi} \right)^T \right] \right\rangle, \label{eq:def-eff-elast-ten}
\end{equation}
where we define transpose of the fourth-rank tensor $\bnabla \ten{\chi}$ acting on the first two indices $\left( \bnabla \ten{\chi} \right)^T_{ijkl} = \left( \bnabla \ten{\chi} \right)_{jikl}$. The expression above provides a direct linear link between the micro- and macroscale elasticity tensors through the known porosity $\theta$ and  the pore-scale geometry captured by $\ten{\chi}$ that will be computed
by solving equations (\ref{eq:unit-chi-prob-1}--\ref{eq:unit-chi-prob-2}).
Here, it is assumed that the skeleton elasticity tensor $\ten{C^\scr{sk}}$ is constant in space. 
The brackets  denote the volume average over the interior unit-cell volume, which is filled
either by solid or by fluid
\begin{equation}
\langle f \rangle = \frac{1}{l^3} \int_{V_\sigma} f \,\mathit{dV}, \label{eq:def-vol-avg}
\end{equation}
where $V_{\sigma}$ is the volume of either the solid or fluid phase. In general, the effective properties should be re-evaluated as the porosity changes. However, we assume small displacements of the effective system so that the porosity is roughly constant and therefore it is enough to compute the averages only once. 

%
The microscale displacement tensor $\ten{\chi}$ is defined in the solid phase (having volume
$V_s$) of the unit-cell and is the solution of the following problem,
%
\begin{align}
\bnabla \cdot \left[ \ten{C}^\scr{sk} : \frac{1}{2} \left\{ \bnabla \ten{\chi}
+ \left(\bnabla \ten{\chi} \right)^T \right\} \right] & = 0, \label{eq:unit-chi-prob-1}\\
\left[ \ten{C}^\scr{sk} : \frac{1}{2} \left\{ \bnabla \ten{\chi}
+ \left(\bnabla \ten{\chi} \right)^T \right\} \right] & \cdot \hn = 
\left[ \ten{C}^\scr{sk} : \ten{\delta^{(4)}} \right] \cdot \hn. \label{eq:unit-chi-prob-2}
\end{align}
Here, the fourth-rank identity tensor appearing in the boundary condition is defined as $\left(\ten{\delta^{(4)}}\right)_{ijkl} = \delta_{ik} \delta_{jl}$. This equation corresponds to a standard steady linear-elasticity problem generalised to third-rank tensor subjected to different combinations of surface loading (applied on the interface between the solid skeleton and the pore fluid) in order to characterise the response of the structure to all possible surface loading scenarios. To complete the formulation, periodic boundary conditions are applied to solid surfaces, which are in contact with the boundary  of the unit-cell. In order to render the solution unique,
\refAll{we impose constraints on average values of displacement\cite{mei2010homogenization}
using penalty terms in weak formulation\cite{github2016UgisShervin}.}

The unit-cell domain has been meshed\footnote{We use mesh spacing $\Delta s_1 = 0.025\,l$ at the solid skeleton and $\Delta s_2 = 0.10\,l$ at the boundaries of the unit-cell. We have also carried out the simulation on coarser meshes ($\Delta s_1 = 0.05\,l$ and $\Delta s_2 = 0.20\,l$; $\Delta s_1 = 0.10\,l$ and $\Delta s_2 = 0.40\,l$) and, by observing the convergence of results, we
set entries containing only numerical noise to zero.} using GMSH software\cite{geuzaine2009gmsh} and equations (\ref{eq:unit-chi-prob-1} -- \ref{eq:unit-chi-prob-2}) has been solved using FreeFEM++\cite{MR3043640}. The sphere radius is $R = 0.3\,l$ and the cylinder radius is $r = 0.1\,l$, which results in porosity $\theta = 0.85$. Assuming the Poisson's ratio to be $\nu = 0.3$ (due to the linearity of the problem, the solution is valid for any Young's modulus value $E$), the effective
elasticity tensor for the cubic-symmetric poroelastic medium in
Voigt notation\cite[p. 136]{cowin2013continuum}
is
\begin{equation}
\ten{C} = \left( \begin{array}{cccccc}
4.792 & 0.239 & 0.239 & 0 & 0 & 0 \\
0.239 & 4.794 & 0.239 & 0 & 0 & 0 \\
0.239 & 0.239 & 4.792 & 0 & 0 & 0 \\
0 & 0 & 0 & 0.246 & 0 & 0 \\
0 & 0 & 0 & 0 & 0.246 & 0 \\
0 & 0 & 0 & 0 & 0 & 0.246 \\
\end{array} \right) \cdot 10^{-2} E. \label{eq:eff-elast-result-cubic}
\end{equation}
%
Note that the entries of the effective elasticity tensor are significantly smaller (up to 300 times) compared to these of the skeleton elasticity. This is because most of the solid mass has been removed, leaving only $15\%$ solid volume fraction if compared to a completely filled case, hence the resulting material is much softer. Another observation is that the shear coefficients (diagonal elements in $3 \times 3$ bottom right matrix block) are now much smaller compared  to pressure-wave coefficients\cite[p. 22]{mavko2009rock}
(diagonal elements in $3 \times 3$ top left matrix block). This effect can be attributed to connecting net of cylinders between spheres; the cylinders are much easier to bend, compared to bulk material, i.e. there is no continuous
support from the sides, as there would be in the continuous material case.
The effective tensor can be characterized by three independent parameters\footnote{The difference of $2 \cdot 10^{-5}$ between the first and second second diagonal element is attributed to numerical error that could be caused by, for example, imperfections in the generated mesh.} and has the form of a material with cubic symmetry \cite[p. 99]{cowin2013continuum}.

The other two elasticity-related effective parameters $\ten{\alpha}$ and $\mathcal{E}$ can be computed through expressions\footnote{For explicit relationships obtained using multi-scale expansion, see supplementary appendix~\ref{app:sec-app:subderiv-from-elast}.} involving the inverse of the fourth-rank elasticity tensor effective elasticity $\ten{C}$ \cite[eqs. 2.4,2.7,2.8]{gopinath2011elastohydrodynamics}. Alternatively, one can solve for an additional test displacement field $\vec{\eta}$ governed by the linear elasticity problem
\begin{align}
\bnabla \cdot \left[ \frac{\ten{C}^\scr{sk}}{E} : \frac{1}{2} \left\{ \bnabla \vec{\eta}
+ \left(\bnabla \vec{\eta} \right)^T \right\} \right] & = 0, \label{eq:unit-eta-prob-1}\\
\left[ \frac{\ten{C}^\scr{sk}}{E} : \frac{1}{2} \left\{ \bnabla \vec{\eta}
+ \left(\bnabla \vec{\eta} \right)^T \right\} \right] & \cdot \hn = 
\ten{\delta} \cdot \hn. \label{eq:unit-eta-prob-2}
\end{align}
%
Note that this problem is driven by different surface forcing compared to the $\ten{\chi}$ problem, but is otherwise solved under the same conditions. Given $\vec{\eta}$, the pore-pressure contribution tensor is computed from,
\begin{equation}
\ten{\alpha} = \theta \ten{\delta} + \left\langle \frac{\ten{C}^\scr{sk}}{E} : \frac{1}{2}
\left[ \bnabla \vec{\eta} + \left(\bnabla \vec{\eta}\right)^T \right] \right\rangle,
\label{eq:def-eff-alpha-ten}
\end{equation}
and the coefficient describing the poroelastic material response to the time-variation
of the pore pressure is computed from 
\begin{equation*}
\mathcal{E} = \left\langle \frac{\bnabla \cdot \vec{\eta}}{E} \right\rangle.
\end{equation*}
%
By solving (\ref{eq:unit-eta-prob-1}--\ref{eq:unit-eta-prob-2}) for the particular microstructure with cubic symmetry shown in Fig.~\ref{fig:pemedia-design-def-cubicSym}$b$, we obtain  
\begin{equation*}
\ten{\alpha} = 0.9789\,\ten{\delta} \qquad \textrm{and}\qquad \mathcal{E} = 0.1571\, E^{-1}.
\end{equation*}
The tensor $\ten{\alpha}$ is not unity, therefore one can conclude that the solid
phase of the poroelastic material is not incompressible. This conclusion agrees
with the used elasticity parameter (we use $\nu = 0.3$, while incompressible
solids have $\nu = 0.5$). However, it is non-trivial to use tensor $\ten{\alpha}$
as a measure of compressibility, because it is also a function of
porosity $\theta$.


\subsubsection{Microscale problem for permeability tensor}
Turning attention to the fluid flow, the poroelastic material is characterized
by a permeability \cite{darcy1856fontaines}, which in our framework is obtained by solving the following Stokes problem in the interior unit-cell (Fig.~\ref{fig:pemedia-design-def-cubicSym}b),
\begin{align}
- \bnabla \vec{\mathcal{A}} + \bnabla^2 \ten{\mathcal{K}} & = - \ten{\delta}, \label{eq:unit-perm-prob-1} \\
\bnabla \cdot \ten{\mathcal{K}} & = \vec{0}. \label{eq:unit-perm-prob-2}
\end{align}
Here, $\ten{\mathcal{K}}$ is a second-rank tensor field; $\mathcal{K}_{ij}$ is the $i$th velocity component of $j$th vectorial Stokes problem associated with the pressure field  $\mathcal{A}_j$. The three vectorial Stokes problems characterize how the pore flow responds to volume forcing in one spatial direction at a time. 
This equation system is complemented with no-slip boundary condition at the interface with the solid skeleton, and periodic boundary conditions at the sides of the unit-cell. For uniqueness, we require that the average pressure field is zero. The effective permeability tensor for the cubic symmetric material is then obtained by averaging the  field $\ten{\mathcal{K}}$ over the fluid volume
in the unit-cell as
\begin{equation}
\ten{K} = \langle \ten{\mathcal{K}} \rangle = 2.32 \cdot 10^{-2}\, l^2\,\ten{\delta}.
\label{eq:interior-K-res-cubic}
\end{equation}
We observe that the permeability tensor is characterized by 
a single constant value,
which is characteristic for isotropic flow in the poroelastic medium.
Hence, the cubic-symmetry of the poroelastic material is not visible in permeability
tensor. 

\subsubsection{Effects of significant anisotropy at the pore-scale} \label{sec:monoc-sym-mat}

\begin{figure}
  \begin{center}
  \includegraphics[width=1.0\linewidth]{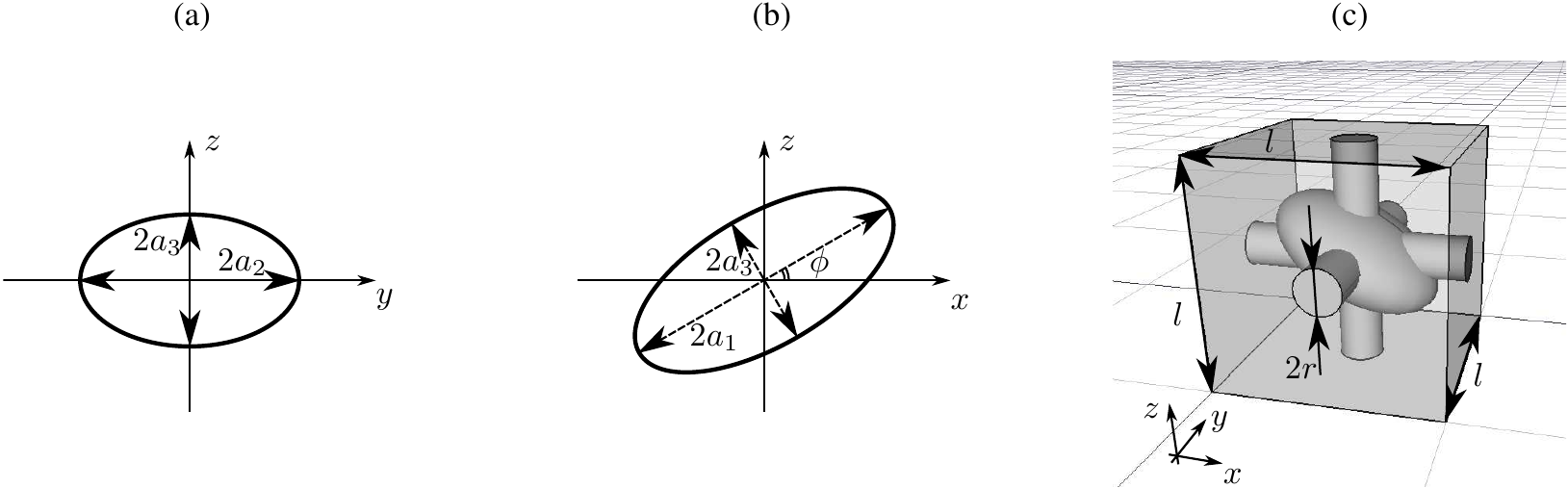} 
  \end{center}
  \caption{
Definition of an ellipsoid in the center of representative volume element
in poroelastic medium. Panel (a) shows the ellipsoid slice in $(y,z)$ plane.
This particular ellipsoid is aligned with axis. Panel (b) shows the tilt of the
ellipsoid around $y$ axis with angle $\phi$. Panel (c) shows the
micro-structure geometry with one plane of symmetry (monoclinic symmetric
material). The ellipsoid parameters
are $a_1 = 0.4\, l$, $a_2 = 0.3\, l$ and turn angle is $\phi = 30^{\circ}$. Cylinder
radius is $r = 0.1\,l$.  \label{fig:pemedia-design-def-monocSym}}
\end{figure}

With the aim of providing a framework that can cope with any periodic microstructure, we compute the effective parameters for a strongly anisotropic microstructure, namely, a tilted ellipsoid at the center of the unit-cell, as shown in Fig.~\ref{fig:pemedia-design-def-monocSym}. Note that the cubic symmetry in the poroelastic material is broken by this structure, since it has less symmetry planes compared to the cubic unit cell.
The ellipsoid semi-axes have lengths $a_1$, $a_2$ and $a_3$ in $x$, $y$ and $z$ directions,
respectively. The ellipsoid can be rotated around the $y$ axis by an angle
$\phi$ as shown in Fig.~\ref{fig:pemedia-design-def-monocSym}b. In order to
assess how  anisotropy at the pore-scale modifies the effective parameters at the macroscale, we choose parameters that break as many symmetries as possible; $a_1 = 0.4\, l$, $a_2 = 0.3\, l$ and $a_3 = 0.2\, l$ and  $\phi = 30^{\circ}$. This results in a porosity $\theta = 0.86$ and structure
with only one symmetry plane.

Using the expression  (\ref{eq:def-eff-elast-ten}), we evaluate the effective elasticity tensor for the titled ellipse structure by solving the microscale displacement problem for $\ten{\chi}$ (\ref{eq:unit-chi-prob-1}-- \ref{eq:unit-chi-prob-2}),
\begin{equation}
\ten{C} = \left( \begin{array}{cccccc}
4.668 & 0.283 & 0.233 & 0 & 0.139 & 0 \\
0.283 & 4.638 & 0.209 & 0 &-0.002 & 0 \\
0.233 & 0.209 & 4.031 & 0 & 0.084 & 0 \\
0 & 0 & 0 & 0.168 & 0 & -0.010 \\
0.139 &-0.002 & 0.084 & 0 & 0.205 & 0 \\
0 & 0 & 0 & -0.010 & 0 & 0.258 \\
\end{array} \right) \cdot 10^{-2}\,E. \label{eq:eff-elast-result-monoc}
\end{equation}
The form of $\ten{C}$ correponds to a  monoclinic material symmetry \citep[p. 96]{cowin2013continuum}, and nearly all the elements of the tensor differ from each other, as expected due to the anisotropic pore-structure. 
This symmetry property is invariant with respect to transformation of coordinate axes; by using a different coordinate system, one would only be able to reposition the zero entries in the elasticity tensor to different rows or columns. Moreover, the magnitude of the elements are of the same order as for the cubic-symmetric geometry, since the two cases have nearly the same porosity. 

By solving for the displacement field $\vec{\eta}$ problem
(\ref{eq:unit-eta-prob-1}--\ref{eq:unit-eta-prob-2}), we obtain via expression (\ref{eq:def-eff-alpha-ten}), 
\begin{equation*}
\ten{\alpha} = \left( \begin{array}{ccc}
0.9793 & 0 & -0.0009  \\
0 & 0.9795 & 0  \\
-0.0009 & 0 & 0.9821  \\
\end{array} \right),
\end{equation*}
where we observe that the tensor cannot be characterized by a single constant
as for the cubic-symmetric material, but by four distinct coefficients.
The tensor $\ten{\alpha}$ can be understood as a measure of volume force
in the poroelastic medium
caused by pore-pressure gradient via seepage flow.
The diagonal terms are different because of the different lengths of the
ellipsoid semi-axes $a_1$, $a_2$ and $a_3$.
Due to this difference,
the area of the solid structure (see Fig.~\ref{fig:pemedia-design-def-monocSym}$c$),
exposed to flow in $x$, $y$ and $z$ directions, is different. Consequently,
it results in a different drag force for the same pore-pressure gradient
in different directions.
The off-diagonal term is, on the other hand, caused by the tilt
of the ellipsoid, which gives raise to non-zero pressure force, when projected in 
$x$ direction, due to flow in $z$ direction and vice-versa.
The elastic response to time variation of pressure is
\begin{equation*}
\mathcal{E} = 0.1465\, E^{-1}.
\end{equation*}
The slightly smaller value for the ellipsoid (compared to the sphere) can be attributed to the shape of the pore-structure at the center of the unit-cell. The ellipsoid is thinner in one direction compared to sphere, which leads to larger strain from the same displacement values and consequently a body which is less compressible.

By solving the fluid problem (\ref{eq:unit-perm-prob-1}--\ref{eq:unit-perm-prob-2})
at the pore-scale for the monoclinic symmetric
geometry we obtain effective permeability tensor
\begin{equation}
\ten{K} = \left( \begin{array}{ccc}
2.51 & 0 &-0.10 \\
0 & 2.25 & 0 \\
-0.10 & 0 & 2.22	
\end{array} 
\right) \cdot 10^{-2}\,l^2. \label{eq:interior-K-res-monoc}
\end{equation}
The second-rank permeability tensor has a similar features as the pore-pressure
contribution tensor $\ten{\alpha}$. It has four independent parameters, which are
again set by four geometric parameters -- semi-axes lengths $a_1$, $a_2$ and $a_3$ as
well as turn angle $\phi$. The off-diagonal term shows that a tilted ellipse
generates flow in the $x$ direction if exposed to pressure a gradient along $z$,
and vice-versa. This effect is the same as for a tilted plate exposed to
an incoming parallel free stream. Due to the tilt, a tangential flow
with respect to the incoming free stream appears.

\subsection{Microscale problems for  interface conditions} \label{sec:sub:bc-eq-res}
In order to finalize the homogenized model, one needs effective tensors for
the velocity boundary conditions of the free fluid in contact
with the poroelastic material.
We can determine the velocity boundary conditions (the interface permeability $\ten{K}^\scr{if}$ and the Navier-slip tensor $\ten{L}$) by solving a set of microscale problems in an interface-cell (Fig.~\ref{fig:pemedia-design-def-cubicSym}$a$) as shown for rigid porous media by L\={a}cis \& Bagheri\cite{lacis2016framework}.

To derive the microscale problems in the interface unit cell, we decompose the flow above
the interface into a fast flow $\vec{U}$ and a perturbation $\vec{u}^+$ flow 
\begin{equation}
\vec{u} = \vec{U} + \vec{u}^+. \label{eq:fast-slow-decomp}
\end{equation}
Below the interface there is only slow flow $\vec{u}^-$.
The flow perturbation velocity is the cause of the slip velocity at the interface
with porous or poroelastic material\cite{lacis2016framework}. Note that the global 
pressure difference $\Delta P$
is driving the fast flow $\vec{U}$ above the
interface as well as the slow flow $\vec{u}^-$ below the interface, whereas
$\vec{u}^+$ is driven by the processes
in the poroelastic material. In order to
determine the perturbation velocity, one can arrive to a solution in
the interface-cell (Fig.~\ref{fig:pemedia-design-def-cubicSym}$a$)
using an assumption of linear superposition
\begin{equation*}
\vec{u}^{\pm} = \pdt \vec{v} - \frac{\ten{\mathcal{K}}^{\pm}}{\mu} \cdot
\bnabla p^{-} + \ten{\mathcal{L}}^{\pm} : \left[ \bnabla \vec{u} +
\left( \bnabla \vec{u} \right)^T \right],
\end{equation*}
in which the unknown tensorial fields $\ten{\mathcal{K}}^{\pm}$ and
$\ten{\mathcal{L}}^{\pm}$ are velocity response fields to test forcing,
which corresponds to pressure gradient (volume forcing) and surface
stress (surface forcing).
Fields $\ten{\mathcal{K}}^{-}$ and $\ten{\mathcal{L}}^{-}$ are correspondingly
defined below the interface (illustrated using the dash red line), and fields
$\ten{\mathcal{K}}^{+}$ and $\ten{\mathcal{L}}^{+}$ are defined above the interface,
as sketched in
Fig.~\ref{fig:illustr-interface-cell}.
In the next sections, we introduce problems for all of the introduced test fields
and solve them for both microscale geometries.



\subsubsection{Interface permeability tensor $\ten{K}^\scr{if}$} \label{sec:subsub:bc-eq-res-K}
\begin{figure}
  \begin{center}
  \includegraphics[width=0.8\linewidth]{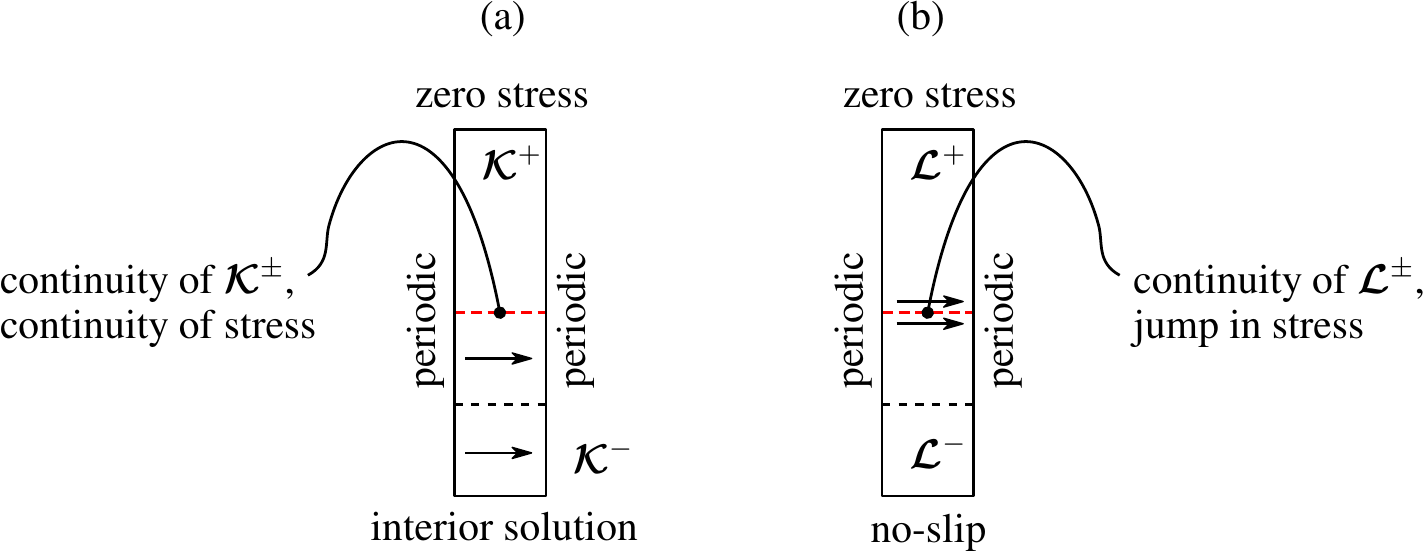} 
  \end{center}
  \caption{
A two-dimensional illustration of the interface-cell problem for $\mathcal{K}_{ij}$
(a) and $\mathcal{L}_{ijk}$ (b). The interface-cell has volume forcing
in the $x$ direction below
the interface (red dashed line) for the $\mathcal{K}_{i1}$ problem (a). The interface 
cell has a jump in stress at the interface location, which results in a force
along $x$ for the $\mathcal{L}_{i13}$ problem (b). The sides of the interface-cell
are exposed to periodic boundary conditions, and the top -- to zero-stress
condition. At the bottom of the cell either the
interior solution (for the $\mathcal{K}_{ij}$ problem)
or no-slip (for the $\mathcal{L}_{ijk}$ problem) is prescribed.
The dashed lines correspond
to boundaries of unit-cells. Above the last unit cell, there is
only free-fluid.
\label{fig:illustr-interface-cell}}
\end{figure}

The interface permeability second-rank tensor field below the interface $\ten{\mathcal{K}}^{-}$ is governed by 
\begin{align}
- \bnabla \vec{\mathcal{A}}^{-} + \bnabla^2 \ten{\mathcal{K}}^{-} & = - \ten{\delta}. \label{eq:intf-unit-perm-prob-1}
\end{align}
The boundary conditions
at the sides of this domain are periodic (same as for interior cell),
however, at the bottom of the domain one has to use the interior solution
provided by the interior problem (\ref{eq:unit-perm-prob-1}--
\ref{eq:unit-perm-prob-2}), 
see Fig.~\ref{fig:illustr-interface-cell}$a$.
At the interface,
the continuity of permeability fields,
\begin{equation}
\ten{\mathcal{K}}^{-} = \ten{\mathcal{K}}^{+},
\end{equation}
is employed.
Since the permeability field in the
upper part of the interface-cell is also unknown, it can be found using
an unforced Stokes momentum equation:
\begin{align}
- \bnabla \vec{\mathcal{A}}^{+} + \bnabla^2 \ten{\mathcal{K}}^{+} & = 0. \label{eq:intf-unit-perm-prob-3}
\end{align}
The volume force in Stokes equations
above the interface does not exist, because the global driving
pressure above the interface acts on the fast flow, not on the
perturbation (\ref{eq:fast-slow-decomp}).
To complete the two-domain formulation
of the interface-cell, incompressibility constraints are added,
no-slip condition is imposed at the surface of solid skeleton,
periodic boundary conditions are applied at the sides,
zero stress condition at the top of the cell,
and stress continuity at the interface,
\begin{equation}
\left\{ -\ten{\delta} \vec{\mathcal{A}}^{-} + \left[ \bnabla \ten{\mathcal{K}}^{-}
+ \left( \bnabla \ten{\mathcal{K}}^{-} \right)^T \right] \right\} \cdot \hn = 
\left\{ -\ten{\delta} \vec{\mathcal{A}}^{+} + \left[ \bnabla \ten{\mathcal{K}}^{+}
+ \left( \bnabla \ten{\mathcal{K}}^{+} \right)^T \right] \right\} \cdot \hn,
\label{eq:intf-unit-perm-prob-bc-K}
\end{equation}
as shown in Fig.~\ref{fig:illustr-interface-cell}$a$.
One can observe from equations
(\ref{eq:intf-unit-perm-prob-1}--\ref{eq:intf-unit-perm-prob-bc-K}) that
the outlined problem is a combination of Stokes problems exposed to unit
volume-forcing in all possible directions below the interface.
For example, the first column of the tensor
$\ten{\mathcal{K}^{\pm}}$ field ($\mathcal{K}^{\pm}_{i1}$) corresponds
to flow response to unit volume forcing in the $x$ direction,
as shown in Fig.~\ref{fig:illustr-interface-cell}$a$.
Thus, this test
problem is used to characterize the pore-pressure gradient (which is a volume
force) contribution to velocity near the interface.
Finally, after solving the coupled two-domain Stokes problem, one can obtain the
effective interface permeability, by employing volume average $\ten{K}^\scr{if} = 
\langle \ten{\mathcal{K}}^{+} \rangle$ in $l^3$ cube above the interface. 

Evaluating\footnote{Similarly
as in interior simulations, we carry out investigations
using three different resolutions, $\Delta s_1 = 0.05\,l$
and $\Delta s_2 = 0.20\,l$, $\Delta s_1 = 0.0625\,l$
and $\Delta s_2 = 0.2500\,l$ and $\Delta s_1 = 0.10\,l$
and $\Delta s_2 = 0.40\,l$. The interface-cell is chosen to encompass
three structures below the interface and is extending for the same
distance in the free fluid.} the interface permeability tensor gives
\begin{equation}
\ten{K}^\scr{if} = \left( \begin{array}{ccc}
3.20 & 0 & 0  \\
0 & 3.27 & 0  \\
0 & 0 & 2.33  \\
\end{array} \right) \cdot 10^{-2}\,l^2 \ \ \mbox{and} \ \ 
\ten{K}^\scr{if} = \left( \begin{array}{ccc}
3.42 & 0 & 0  \\
0 & 3.21 & 0  \\
-0.10 & 0 & 2.23  \\
\end{array} \right) \cdot 10^{-2}\,l^2, \label{eq:intf-K-results}
\end{equation}
for cubic-symmetric and monoclinic-symmetric pore-scale geometries, respectively.
One can observe that the interface-normal permeability components (for velocity
component $u_z$, last row of the matrix)
are the same as interior ones (set by conservation of
mass\cite{lacis2016framework}), but
tangential components are different. The interface permeability for the tangential
components is larger than the interior one, which can be explained by the
fact that there is no friction from the upper part of the interface-cell, while
the interior cell would be exposed to friction from neighbouring cells. In addition,
the interface permeability matrix for monoclinic symmetric pore geometry
is no more symmetric, that is, the effect of
anisotropy has vanished for $u_x$ velocity component at the interface. The reason
for this coefficient vanishing could be the fact that
the anisotropy is too weak to overcome dissipation at the cylinder on the top of
the last micro-structure and does not contribute for the interface velocity.

\subsubsection{Interface Navier-slip tensor $\ten{L}$} \label{sec:subsub:bc-eq-res-L}
When there is a flow over a porous or poroelastic material, there is
a slip velocity induced proportional to free fluid shear stress,
as theoretically derived by Mikeli{\'c} \&  J{\"a}ger\cite{mikelic2000interface}
for a one-dimensional channel flow and
by L\={a}cis \& Bagheri\cite{lacis2016framework} for
a general three-dimensional set-up. 

The slip length problem, unlike the permeability problem, is specific
to the interface, i.e. there exists no analogous problem for the interior domain.
The slip third-rank tensor fields are also governed
by Stokes  equations
\begin{align}
- \bnabla \ten{\mathcal{B}}^{-} + \bnabla^2 \ten{\mathcal{L}}^{-} & = 0, \label{eq:intf-unit-perm-prob-5} \\
- \bnabla \ten{\mathcal{B}}^{+} + \bnabla^2 \ten{\mathcal{L}}^{+} & = 0, \label{eq:intf-unit-perm-prob-6}
\end{align}
below and above the interface, respectively, see
Fig.~\ref{fig:illustr-interface-cell}$b$.
As before, the fields in two
domains are connected through continuity
condition $\ten{\mathcal{L}}^{-} = \ten{\mathcal{L}}^{+}$. Momentum equations
in this test problem are not exposed to volume test forcing. The only non-triviality
in this coupled two-domain problem is jump in stresses across the interface
\begin{equation}
\left\{ -\ten{\delta} \ten{\mathcal{B}}^{-} + \left[ \bnabla \ten{\mathcal{L}}^{-}
+ \left( \bnabla \ten{\mathcal{L}}^{-} \right)^T \right] \right\} \cdot \hn = 
\left\{ -\ten{\delta} \ten{\mathcal{B}}^{+} + \left[ \bnabla \ten{\mathcal{L}}^{+}
+ \left( \bnabla \ten{\mathcal{L}}^{+} \right)^T \right] \right\} \cdot \hn
+ \ten{\mathcal{J}}, \label{eq:intf-unit-perm-prob-bc-L}
\end{equation}
where the third-rank interface stress jump tensor is defined
as $\left(\ten{\mathcal{J}}\right)_{ikl} = \delta_{ik} n_l$. At the sides of the interface
cell we employ periodic boundary conditions, and at the top of the interface
cell the zero-stress boundary condition is used, \refAll{and at the
solid structure we have no-slip condition}. At the bottom of the interface-cell,
however, the no-slip condition is used, because there
is no $\ten{\mathcal{L}}$ field
in the interior.
By looking at the governing equations of the test problem
(\ref{eq:intf-unit-perm-prob-5}--\ref{eq:intf-unit-perm-prob-bc-L}) one can
conclude that this test problem is exposed to test forcing at the interface.
For example, the problem with second and third tensor indices being $j = 1$ and
$k = 3$ ($\mathcal{L}_{i13}$) corresponds to unit forcing in the $x$ direction
at the interface, as shown in Fig.~\ref{fig:illustr-interface-cell}$b$.
Thus this test problem is used to characterize the free-fluid surface stress
contribution to the interface velocity.
Finally, after solving the coupled two-domain Stokes problem, one can obtain the
effective interface slip length, by employing the same volume
average $\ten{L} = 
\langle \ten{\mathcal{L}}^{+} \rangle$ above the interface. 

Evaluating the slip length tensor at the interface gives
\begin{equation}
\ten{L} \cdot \hat{z} = \left( \begin{array}{ccc}
0.183 & 0 & 0  \\
0 & 0.187 & 0  \\
0 & 0 & 0  \\
\end{array} \right)\,l \qquad \mbox{and} \qquad
\ten{L} \cdot \hat{z} = \left( \begin{array}{ccc}
0.191 & 0 & 0  \\
0 & 0.186 & 0  \\
0 & 0 & 0  \\
\end{array} \right)\,l, \label{eq:intf-L-results}
\end{equation}
for cubic-symmetric and monoclinic-symmetric pore-scale geometries, respectively.
In the planar interface case, the only meaningful interface problems for the
Navier-slip length are those which relate slip length with velocity
shear in interface-normal direction\cite{lacis2016framework},
therefore we have presented the
slip length tensor dot product with unit normal vector $\hat{z}$ normal to the interface
with poroelastic material (other entries in this third-rank tensor
are zero).
The non-zero entries in both
matrices corresponds to influence on velocity components $u_x$ and
$u_y$ from velocity strains $\left(\pd_z u_{x} + \pd_x u_{z} \right)$ and
$\left(\pd_z u_{y} + \pd_y u_{z} \right)$, respectively.
For the cubic-symmetric
case, both coefficients are similar due to cubic symmetry of the pore-scale
geometry. That is, the structure (see Fig.~\ref{fig:pemedia-design-def-cubicSym}$b$)
is the same in the $x$ and the $y$ directions.
The monoclinic-symmetric structure, on the other hand, is different
in the $x$ and the $y$ directions (see Fig.~\ref{fig:pemedia-design-def-monocSym}$c$).
However, the slip-length coefficients are still similar, because (i) the
top cylinder acts exactly the same way in both $x$ and $y$ directions and (ii) the tilted
ellipsoid slows down the velocity in both $x$ and $y$ directions similarly in
integral sense.
For $u_z$ component, there is no contribution in both cases, because
the interface-normal penetration velocity is governed by mass conservation
alone.

\section{Poroelastic material response to free fluid vortex above it} \label{sec:results-cavity}

The purpose of this section is to exemplify the proposed numerical framework by illustrating the response of the  cubic-symmetric and monoclinic-symmetric poroelastic materials from section \ref{sec:effective-prop}  to free fluid vortex above it. 
In order to create a two-dimensional fluid vortex, we consider a steady low-Reynolds number lid-driven cavity problem,  consisting of a free fluid domain $\Omega_f$ and a poroelastic domain $\Omega_p$.
We will validate and characterize the accuracy of the effective continuum description  by comparing to a second approach, in which the whole domain is meshed and the flow field as well as the displacement field are resolved at all spatial scales. 
For the fully resolved numerical studies to be feasible, we do not deform the computational mesh when the microstructure is displaced, which sets an upper limit of displacement we can consider
to roughly $\vec{v} \lesssim 0.1\,l$. Note that this means we consider a one-way interaction problem, that is, the material elasticity does not influence the free fluid, whereas the free-fluid does induce a displacement of the material. 
%
%
The results obtained through this simplification provides new fundamental insight into the physics at the interface between the free-fluid and porous region. It allows us to study how much of the shear stress from the free fluid results in a stronger pore flow and how much of it is borne by the solid.
The theory in sections~\ref{sec:homog-eq} and \ref{sec:effective-prop} and the numerical
implementation of effective equations\cite{github2016UgisShervin} is valid for two-way coupled problems, but the comparison with fully resolved simulations of such systems we leave for future work.

\begin{figure}
  \begin{center}
  \includegraphics[width=0.8\linewidth]{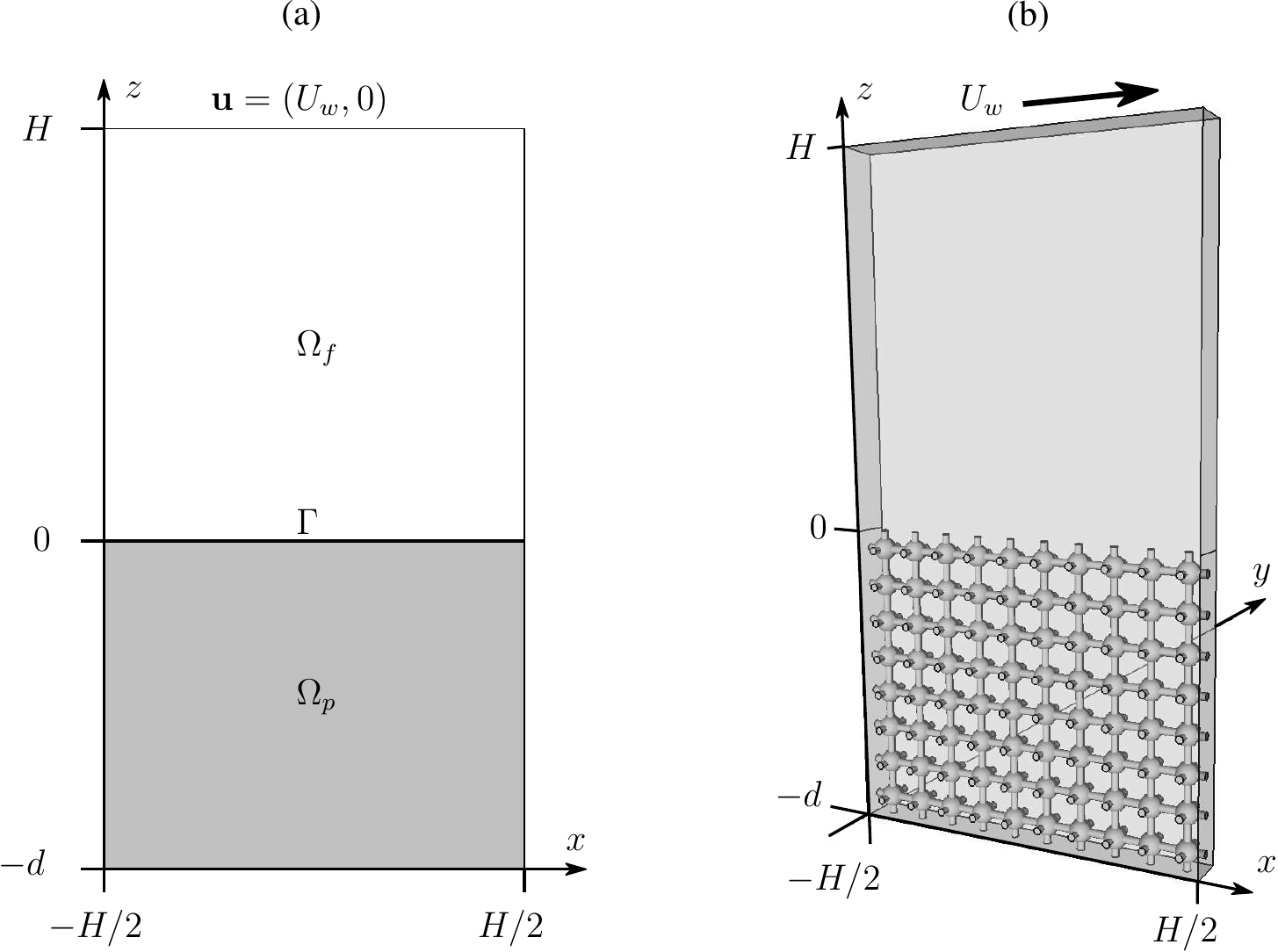} 
  \end{center}
  \caption{
Left frame (a) shows a two-domain -- consisting of free fluid domain $\Omega_f$
and poroelastic domain $\Omega_p$ -- averaged description
of the lid-driven cavity flow in two dimensions.
Right frame (b) shows the lid-driven cavity problem in a three-dimensional
setting, which we use for model validation. It is assumed
that the flow is periodic in the $y$ direction over a length
$l$ (one micro-structure). Cubic-symmetric pore-scale geometry
is used in this drawing. \label{fig:cavity-design-DNS-MOD}}
\end{figure}

\subsection{Effective continuum description} \label{sec:sub:results-effective}
As shown in Fig.~\ref{fig:cavity-design-DNS-MOD}$a$, the two-dimensional (2D) lid-driven cavity has a depth of $H+d$, a length of $H$ and is infinitely wide. The poroelastic medium is confined to $z \leq 0$ and $-H/2 \leq x \leq H/2$. The flow is driven by the top-wall, which moves in the $x$ direction with speed $U_w$.
%
In $\Omega_p$, we solve the effective equations for a poroelastic medium (\ref{eq:dim-pore-momentum}, \ref{eq:dim-pore-pressure}) with the boundary conditions at the wall (\ref{eq:homog-bc-walls}) and at the interface with the free fluid (\ref{eq:homog-p-bc-if},\ref{eq:homog-v-bc-if}). For the free fluid in $\Omega_f$, we solve equations (\ref{eq:NStokes-ms-dimensional-1},
\ref{eq:NStokes-ms-dimensional-2}) with velocity boundary condition at the interface with the poroelastic material (\ref{eq:dim-vel-if-condition}, \ref{eq:dim-vel-if-condition-normal}). At the side and top walls of the cavity, we impose
\begin{equation}
\vec{u} = 0 \ \ \textrm{ at } \ \  x = \pm \frac{H}{2}, \qquad
\vec{u} = \left( U_w, 0 \right) \ \  \textrm{ at } \ \  z = H,
\label{eq:free-fluid-wall-bc}
\end{equation}
respectively.
%
%
The governing equations are discretized using finite-element-method (FEM).
The simulation domain along with equations is defined
in the FreeFEM++\cite{MR3043640} software using
uniform mesh
spacing\footnote{We have investigated results using half of the mesh spacing
$\Delta s = 0.10\,l$ and observed that the slip velocity changed by
$0.4\%$ and the horizontal displacement near the interface changed
by $0.8\%$ at the center of the cavity.}
$\Delta s = 0.20\,l$. We
implement the weak formulation
and solve the time dependent problem in a fully implicit manner, where boundary
conditions at the interface are enforced using Lagrange multipliers\footnote{Exact
weak form can be found in documentation
of the open-source software\cite{github2016UgisShervin}.}.
The effective coefficients, such as elasticity tensor and permeability matrix, are
taken from
section~\ref{sec:effective-prop}.
To render the validation with the resolved model feasible, we select a moderate
scale separation parameter $l/H = 0.1$.

\begin{figure}
  \begin{center}
  \includegraphics[width=1.0\linewidth]{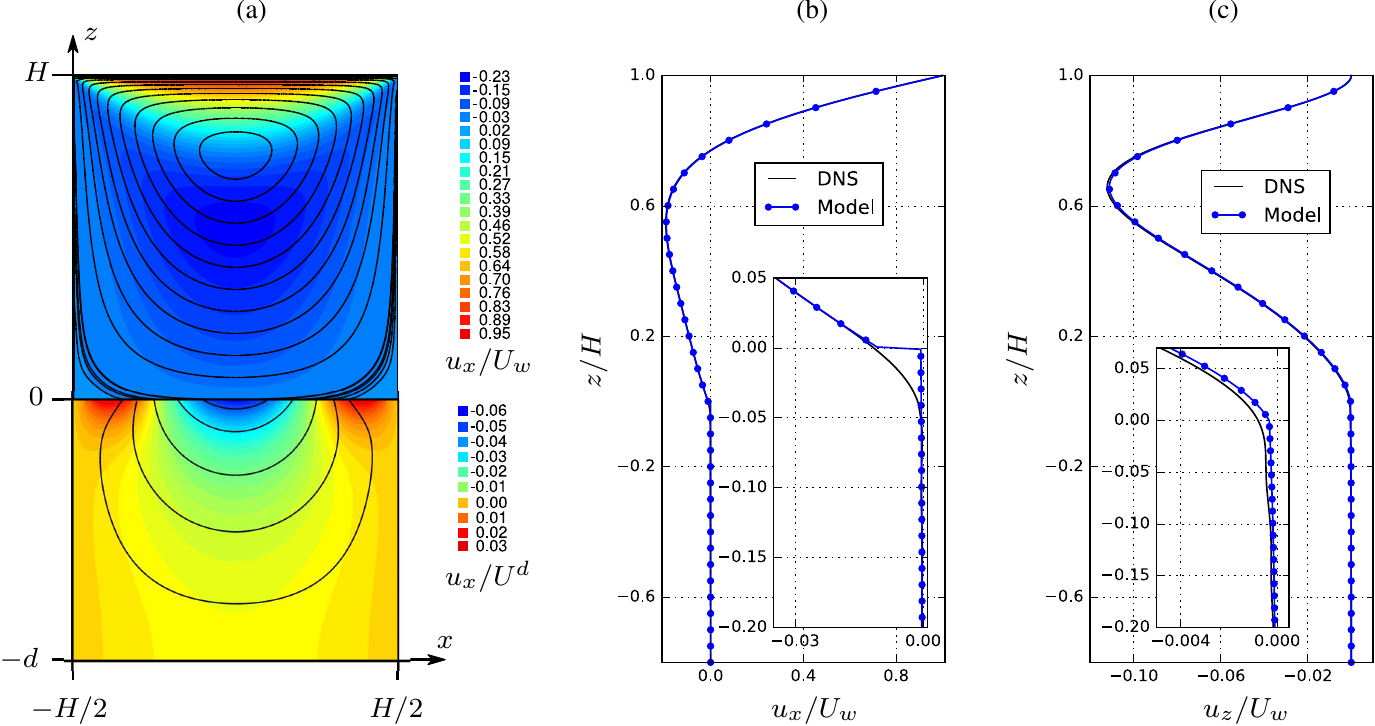} 
  \end{center}
  \caption{
Comparison of flow velocities between effective model of the cavity
and the direct numerical simulation of cubic-symmetric pore-scale geometry.
In the left frame (a) we show model results; the coloured iso-contours corresponds
to stream-wise velocity distribution in free fluid and poroelastic material
and black lines are flow streamlines. \refAll{The stream-wise velocity in the
poroelastic material is normalised with Darcy's velocity $U^d$ and in the
free fluid it is normalized with upper-wall velocity $U_w = 100\,U^d$.}
In the middle frame (b) we show the stream-wise velocity variation
over the vertical coordinate at $x = 0.1\,H$. In the right frame (c) we show
the interface-normal velocity variation over the vertical coordinate at $x = 0.1\,H$.
The insets in frames (b) and (c) show flow results near the interface.
\label{fig:DNS-results-geomA-compMOD}}
\end{figure}

\begin{figure}
  \begin{center}
  \includegraphics[width=1.0\linewidth]{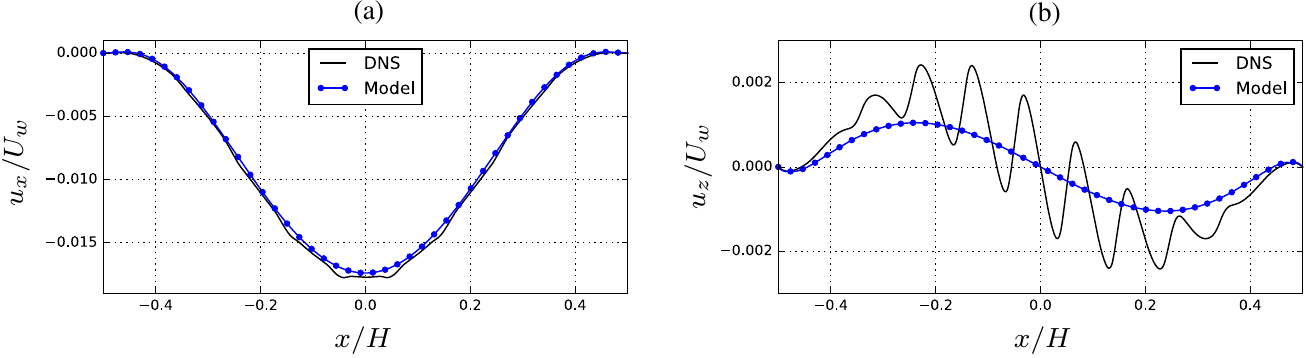} 
  \end{center}
  \caption{
Comparison of flow velocities between effective model of the cavity
and the direct numerical simulation of cubic-symmetric pore-scale geometry near
the interface.
In the left frame (a) we show slip velocity in the cavity problem
near the tip of the solid skeleton at $z = 0.01\,H$. In the right frame
(b) we show the penetration velocity at the same coordinate $z = 0.01\,H$.
\label{fig:DNS-results-geomA-compMOD-intf}}
\end{figure}

\subsubsection{Effective fluid velocity}
We start by presenting the flow field $\vec{u} = \left(u_x, u_z\right)$ in the full domain ($\Omega_p$ and $\Omega_f$)
for the cubic-symmetric poroelastic medium with porosity $\theta = 0.85$.
Streamlines and isocountours of the streamwise fluid velocity are shown in Fig.~\ref{fig:DNS-results-geomA-compMOD}$a$. In the free fluid domain we normalize the results
using the top wall velocity $U_w$ (in the simulations we have set $U_w = 100\ U^d$),
whereas in the poroelastic domain we normalize the
results using Darcy's velocity $U^d \equiv l^2 \Delta P / \left(\mu H \right)$.
The moving top wall creates a circulation in the cavity, where the flow for $x>0$ is directed downwards, and upwards for $x<0$. We observe that due to the vortex in the free fluid, there exists a  transfer of mass and momentum across the interface. This is characterized in more detail from the slip and infiltration (penetration) velocities very close to the interface ($z=0.01H$) in Figs.~\ref{fig:DNS-results-geomA-compMOD-intf}$a$~and ~\ref{fig:DNS-results-geomA-compMOD-intf}$b$, respectively.
The streamwise slip velocity  (Fig.~\ref{fig:DNS-results-geomA-compMOD-intf}$a$) has  a parabolic shape near the poroelastic medium with its minimum (largest magnitude) velocity at the center of the cavity. This velocity component is mainly created by the shear of the free fluid, which is the strongest at $x=0$, before it gradually decays when approaching the sides of the cavity.
The penetration velocity (Fig.~\ref{fig:DNS-results-geomA-compMOD-intf}$b$) shows a macroscopic 
behaviour similar to a sine wave. We observe that for $x > 0$, there is a net mass/momentum transport from free fluid region to the poroelastic region, whereas for $x < 0$ the net mass flow is in the opposite direction. 

Fig.~\ref{fig:DNS-results-geomA-compMOD}$a$ shows that the flow inside the bed circulates in the same direction as in the free-fluid region.
Note that the streamwise velocities in the two
domains differ by roughly three orders of magnitude.
This gives the impression of a clear discontinuity of streamwise velocity component; 
Figs.~\ref{fig:DNS-results-geomA-compMOD}$b$
and \ref{fig:DNS-results-geomA-compMOD}$c$ shows the streamwise $u_x$ and wall-normal $u_z$ velocity profiles for the fixed streamwise position $x = 0.1H$. For the streamwise component (see inset of Fig.~\ref{fig:DNS-results-geomA-compMOD}$b$), one can observe that there is a very slow Darcy flow inside the poroelastic material, whereas near the interface there is a jump to the fast slip velocity of the free flow. The leading-order effective equation for flow inside a poroelastic material, which is  relative Darcy's law (\ref{eq:dim-pore-momentum}), contains only the pressure contribution from the flow and no viscous fluid stress. As a consequence, the fluid inside the poroelastic layer can only be driven by the normal stress (can not respond to shear stress), which  is the reason why the velocity jump arises. In Fig.~\ref{fig:DNS-results-geomA-compMOD}$c$, we see that wall-normal component $u_z$ is continuous across the interface, which follows from mass conservation.
 
We compared the homogenized model results between cubic-symmetric poroelastic material ($\theta = 0.85$) and monoclinic-symmetric poroelastic material ($\theta = 0.86$) and did not observe any significant difference in flow velocities, despite the different levels of anisotropy in the pore microgeometries.
The reason for this outcome is the fact that the introduced anisotropy
results only in higher order corrections of permeability
tensors (\ref{eq:interior-K-res-cubic},\ref{eq:interior-K-res-monoc},
\ref{eq:intf-K-results})
and Navier-slip tensors (\ref{eq:intf-L-results}).


\subsubsection{Effective solid displacement} \label{sec:subsub:results-effective-displ}
\begin{figure}
  \begin{center}
  \includegraphics[width=1.0\linewidth]{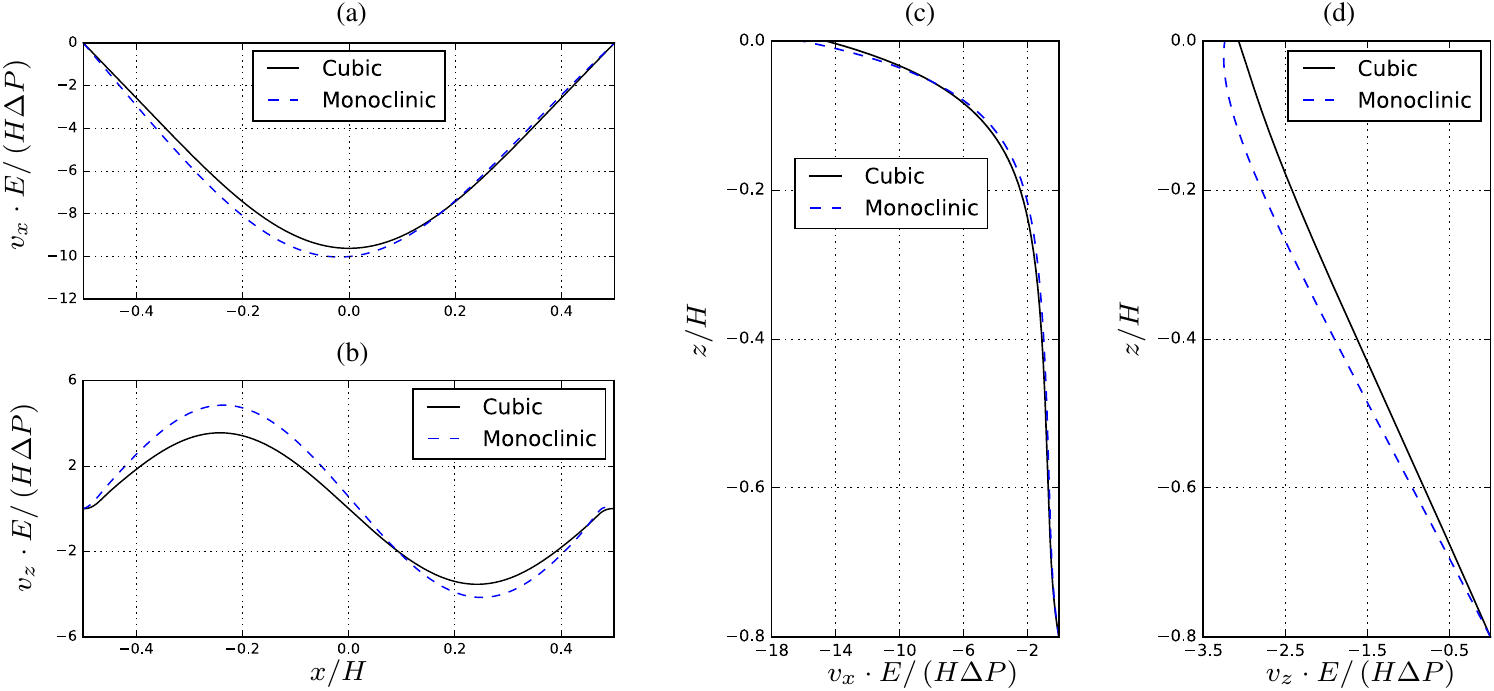}
  \end{center}
  \caption{
Comparison between effective
solid displacement data for cavity
with poroelastic bed
built from cubic-symmetric and monoclinic-symmetric geometries.
In the top left frame (a) we show horizontal displacement
near the tip of the solid skeleton at $z =-0.05\,H$. In the bottom left frame
(b) we show the vertical displacement at the same coordinate $z =-0.05\,H$.
In the middle frame (c) we show the horizontal displacement variation
over the vertical coordinate at $x = 0.15\,H$. In the right frame (d) we show
the vertical displacement variation over the vertical coordinate at $x = 0.15\,H$.
It is estimated that the displacements have to be $\vec{v} \lesssim 0.1\,l$ in
order for the current implementation to be reliable despite the non-deforming mesh.
\label{fig:geomAC-compMOD-displ}}
\end{figure}

In Figs.~\ref{fig:geomAC-compMOD-displ}$a$ and \ref{fig:geomAC-compMOD-displ}$b$ we show displacement $v_x$ and $v_z$ along $x$-coordinate at a fixed  $z = -0.05\,H$ for both cubic-symmetric and monoclinic-symmetric pore-scale geometries.
The displacement fields are
normalized\footnote{Note that due to the linearity of governing equations, the results are valid for any chosen Young's modulus, as long as the deformation is small enough ($\vec{v} \lesssim 0.1\,l$) to neglect the mesh deformation and Poisson's ratio is $\nu = 0.3$. The dimensional normalization is obtained in appendix~\ref{app:sec:app-eq-assume}.} 
using length scale
$H \Delta P / E$.
Moreover, the  $z$-coordinate is chosen to correspond to the centres of connecting cylinders. In this way, the result can easily be compared to fully micro-resolved simulations (see next section), for which it is not straight-forward to define displacement fields in the fluid part of the pores.
For both microstructure geometries, the horizontal displacement $v_x$ near the interface (Fig.~\ref{fig:geomAC-compMOD-displ}$a$) reveals a similar behaviour as for the slip velocity, i.e., the displacement has parabolic shape with maximum magnitude at the center of the cavity.  The horizontal displacement is in the direction of the slip velocity directly above the interface and is caused by the shear stress induced by the overlying flow vortex in the cavity. The vertical displacement $v_z$ near the interface (Fig.~\ref{fig:geomAC-compMOD-displ}$b$), on the other hand, shows a very similar behaviour to that of the penetration velocity, i.e., the  displacement has a sine-like shape. We thus observe that the solid displacement is complying to the tangential and normal fluid velocities near the interface.

In Figs.~\ref{fig:geomAC-compMOD-displ}$c$ and \ref{fig:geomAC-compMOD-displ}$d$ we show $v_x$ and $v_z$ along the $z$-coordinate at a fixed position $x = 0.15\,H$. For the horizontal component $v_x$ (Fig.~\ref{fig:geomAC-compMOD-displ}$c$) two regions can be identified, where the displacement decreases at different rates. The region down to $z \gtrsim -0.2\,H$, for which the decay is very fast, is determined by the shear stress at the interface. Below $z \lesssim -0.2\,H$, it is the slow Darcy flow that induces the small displacement. 
The vertical displacement (Fig.~\ref{fig:geomAC-compMOD-displ}$d$), on the other hand, is entirely governed by the slow penetration velocity inside the medium or
the interface-normal stress at the interface,
depending on which one is the dominating effect.
%

We note that the horizontal displacement (Fig.~\ref{fig:geomAC-compMOD-displ}$a$) has slightly larger magnitude and is skewed for the monoclinic compared to the cubic material. This difference can be explained by a slightly smaller compression elasticity coefficient $C_{xxxx}$ ($4.668 < 4.792$) and by the existence of an elasticity coefficient $C_{xzxx}$ for the monoclinic material (see section~\ref{sec:effective-prop}\ref{sec:interior-prop}\ref{sec:monoc-sym-mat}),
which
relates the strain of $v_x$ in the $x$ direction with the stress in
the interface-normal direction $z$
on the plane with a fixed $x$ coordinate.
The vertical displacement (Fig.~\ref{fig:geomAC-compMOD-displ}$b$) is also larger in magnitude and skewed for the monoclinic geometry. The larger amplitude can again be explained by the difference in compression elasticity coefficient $C_{zzzz}$ ($4.031 < 4.792$), whereas the skewness can be attributed
to $C_{xxxz}$, which provides a coupling between the interface-normal stress of the free fluid
to strain of the vertical displacement $v_z$ in the $x$ direction. We thus conclude that an anisotropy in the microstructure breaks the symmetric behaviour of the solid displacement in the macroscale due to the additional cross terms (such as $C_{xxxz}$) in the effective elasticity tensor.

\subsection{Validation to fully microscale-resolved simulations}
In order to validate the  effective model, a three-dimensional (3D) lid-driven cavity problem is considered (same depth of $H+d$ and length $H$ as before), where the flow is periodic in $y$ direction over length $l$, which is set by the pore-sctructure. Therefore, it is sufficient to
consider the domain shown in Fig.~\ref{fig:cavity-design-DNS-MOD}$b$
(spanning over one microstructure in $y$ direction)
for direct  numerical simulations (DNS). 
\begin{figure}
  \begin{center}
  \includegraphics[width=1.0\linewidth]{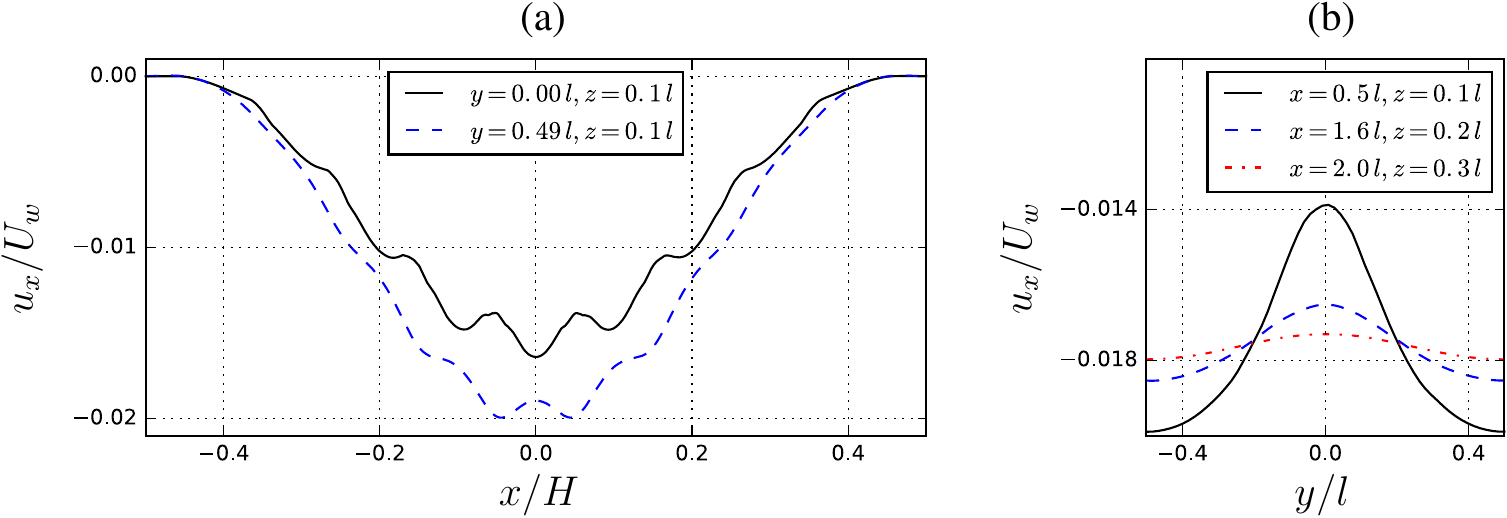} 
  \end{center}
  \caption{
Plots of raw fluid simulation data for cavity with cubic-symmetric poroelastic bed.
In the left frame (a) we show slip velocity in the cavity problem
near the tip of the solid skeleton ($z = 0.01\,H$). We choose to sample the obtained
results at the center of the 3D slice ($y = 0.0\, l$) and near the periodic boundary
($y = 0.49\,l$). In the right frame (b) we show the stream-wise velocity variation
in the periodic direction at points, where the average velocity value is roughly the same.
Coordinates of the line probe are $x = 0.05\,H$, $z = 0.01\,H$; $x = 0.16\,H$, $z = 0.02\,H$ and
$x = 0.20\,H$, $z = 0.03\,H$.\label{fig:DNS-results-line-probes}}
\end{figure}

The domain shown in Fig.~\ref{fig:cavity-design-DNS-MOD}$b$ is defined using
GMSH software \citep{geuzaine2009gmsh} and meshed using spacing $\Delta s_1 = 0.04\,l$
at the solid skeleton and $\Delta s_2 = 0.16\,l$ at the boundaries of the
cavity\footnote{We have carried out same simulations using mesh spacings
$\Delta s_1 = 0.05\,l$ and $\Delta s_2 = 0.20\,l$. The change in fluid slip velocity
was everywhere below $0.3\%$. The change in horizontal solid displacement was everywhere
below $5.5\%$, with maximum at the center of the interface.}.
We import the generated mesh into FreeFEM++ \citep{MR3043640}, in which we define
and solve the governing equations
(\ref{eq:NStokes-ms-dimensional-1}--\ref{eq:solid-ms-dimensional})
with boundary conditions (\ref{eq:solid-fluid-ms-dimensional-bcs},
\ref{eq:free-fluid-wall-bc}). At the walls, we use no-slip velocity for fluid
and zero displacement for solid structure. To simplify the numerical task, we assume a steady flow and neglect inertial effects, which simplifies the Navier-Stokes equations (\ref{eq:NStokes-ms-dimensional-1}--\ref{eq:NStokes-ms-dimensional-2}) to the linear Stokes equations. 
We use Taylor-Hood (P2+P1) elements for the Stokes system, and quadratic elements (P2) for the solid skeleton linear elasticity system (\ref{eq:solid-ms-dimensional}). The resulting linear algebraic system is solved using a GMRES iterative linear solver up to a relative tolerance $e = 10^{-10}$.  For the solid skeleton, we choose the isotropic material that was used as the starting point for the computation of effective properties in section~\ref{sec:effective-prop} (Poisson's ratio $\nu = 0.3$ and Young's modulus $E$ unspecified). We assume that deformations are small enough such that the computational mesh can be kept static;
this limits the range of length scales $\bar{l} = H \Delta P / E$ --
based on Young's modulus $E$,
large scale length $H$ and pressure difference $\Delta P$ -- that
can be considered.
%

\subsubsection{Fully-resolved fluid flow}
In Fig.~\ref{fig:DNS-results-line-probes}$a$, we show the resolved stream-wise velocity (slip velocity)  
near the interface at a distance $z = 0.01\,H$ and coordinates $y = 0$ and $y = 0.49\,l$. The poroelastic bed in this case is built using cubic-symmetric geometry.
We observe that the slip velocity is slightly slower at the center of the cavity slice ($y = 0$) compared to the $xz$-plane near the periodic boundaries ($y = 0.49\,l$). This is because the bulk of the solid material (sphere) is located at the center of each volume element, therefore the surrounding fluid is slowed down more than the fluid near the cylindrical obstructions close to the periodic boundaries.
The stream-wise velocity
distribution in the $y$ direction at three $x$ and $z$ coordinates is shown in Fig.~\ref{fig:DNS-results-line-probes}$b$. 
Moving away from the poroelastic material (increasing $z$ coordinate) leads to a
rapid dissipation of velocity variation; that is, the flow velocity approaches
a constant value with respect to the $y$-coordinate. 
We continue by averaging the DNS quantities in the $y$-direction as,
\begin{equation}
\bar{f}\left(x, z \right) = \frac{ \int\limits_{-l/2}^{l/2} I_d f\left(x, y, z\right)
\,\mathit{dy} }{ \int\limits_{-l/2}^{l/2} I_d
\,\mathit{dy} }, \label{eq:def-DNS-y-avg}
\end{equation}
where $f$ denotes any of $u_x$, $u_y$, $u_z$, $p$, $v_x$, $v_y$ and $v_z$. The function $I_d$ is an indicator function, which is $I_d = 1$ in the domain, where field $f$ is defined, and $I_d = 0$ elsewhere. This average is also known as  intrinsic average\cite{whitaker1998method}, because it is normalized by fluid or solid volumes separately. For convenience, we omit the ``bar'' notation further on.

\begin{figure}
  \begin{center}
  \includegraphics[width=1.0\linewidth]{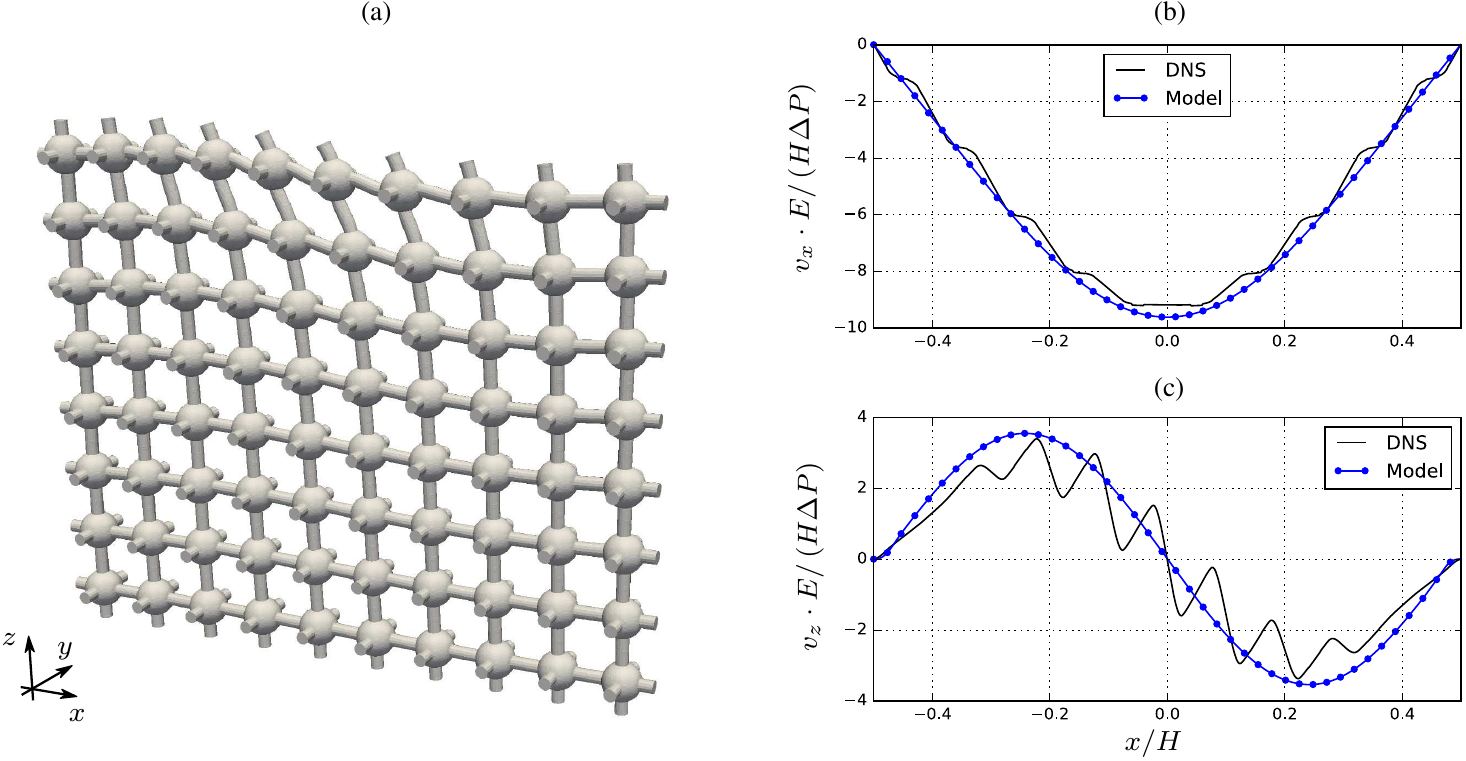} 
  \end{center}
  \caption{
In frame (a) we show a deformed shape of the solid skeleton built
using cubic-symmetric pore-geometry,
exposed to the stress from the surrounding flow. The displacements
are magnified
in order to make the distortion of the skeleton
visible.
In frame (b) we show horizontal displacement
near the tip of the solid skeleton ($z =-0.5\,l$). In frame
(c) we show the vertical displacement at the same coordinate $z =-0.5\,l$.
Both effective model results and fully resolved results are compared. 
\label{fig:DNS-results-solid-displ-lines}}
\end{figure}

The average slip velocity and penetration velocity variations over the $x$ coordinate sampled at $z = 0.01\,H$ are shown in Figs.~\ref{fig:DNS-results-geomA-compMOD-intf}$a$~and ~\ref{fig:DNS-results-geomA-compMOD-intf}$b$,
respectively, together with the effective model curves; one can conclude that the macroscale model is accurate. Note that the effective model is fully non-empirical without any fitting parameters. The penetration velocity (Fig.~\ref{fig:DNS-results-geomA-compMOD-intf}$b$), shows  micro-scale oscillations over one pore-scale structure in the resolved simulations, which are by construction not captured by the effective equations.
The penetration velocity is somewhat under-predicted by the effective model, which may be an indication of a pressure jump \cite{marciniak2012effective,carraro2013pressure}, not modelled in the current work.

%


\subsubsection{Fully-resolved solid displacement}
Fig.~\ref{fig:DNS-results-solid-displ-lines}$a$ shows 
the deformation of the cubic-symmetric structure
due to the free fluid vortex above.
The figure serves only as an illustration, where the solid structure is displaced after the computation, since our implementation is restricted to static meshes. The solid displacement obtained from DNS is compared to the effective model over a horizontal slice at $z = -0.05\,H = -0.5\,l$  in Figs.~\ref{fig:DNS-results-solid-displ-lines}$b$ and \ref{fig:DNS-results-solid-displ-lines}$c$.  Similarly as for the fluid velocity, we observe that the microscale features of the displacement field are not captured by the leading-order effective model, but that the macroscale behavior is in good agreement.

\begin{figure}
  \begin{center}
  \includegraphics[width=0.8\linewidth]{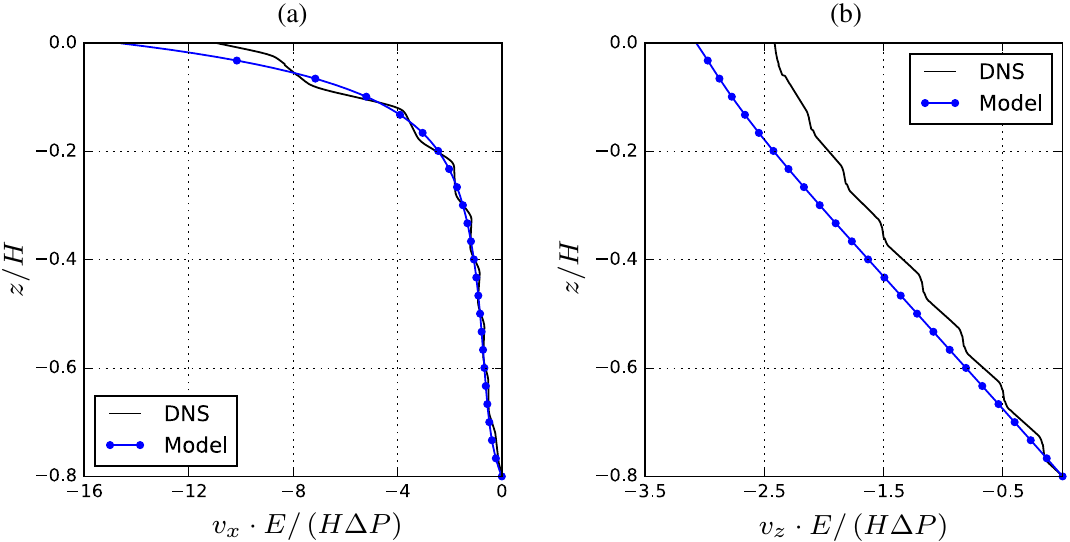} 
  \end{center}
  \caption{
In frame (a) we show horizontal displacement of the poroelastic medium build
from cubic-symmetric pore geometry in depth at coordinate $x = 0.15\,H$. In frame
(b) we show the vertical displacement at the same coordinate $x = 0.15\,H$.
Both homogenized model results and fully resolved results are compared. 
\label{fig:DNS-results-solid-displ-lines-depth}}
\end{figure}

Fig.~\ref{fig:DNS-results-solid-displ-lines-depth}$a$ compares the horizontal displacement along the $z$-coordinate. This shows that the effective model overestimates the horizontal displacement close to the interface,
but is very accurate below $ z \lesssim 0.05\,H$.
This indicates that the inaccuracy is caused by the interface stress boundary condition (\ref{eq:homog-v-bc-if}). The vertical displacement (Fig.~\ref{fig:DNS-results-solid-displ-lines-depth}$b$) is overestimated over the entire depth of the cavity, which suggests that the vertical displacement is mostly governed by the stress at the interface between the free fluid and the poroelastic medium. Despite the inaccuracy at the interface, the model captures the essential qualitative features of the displacement field, including the two different regions of decay of the horizontal displacement. 
The seemingly good agreement of the horizontal displacement in
Fig.~\ref{fig:DNS-results-solid-displ-lines}$b$ arises due to the micro-scale
variations of the DNS solution, as seen
in Fig.~\ref{fig:DNS-results-solid-displ-lines-depth}$a$

Our results show that the free-fluid shear stress transfer to the poroelastic material shear stress is a macroscopic phenomenon. The influence of the interface stress
is ranging over a number of pore-structures; the shear region in Fig.~\ref{fig:DNS-results-solid-displ-lines-depth}$a$ extends over distance of $z \approx 0.3\,H$ containing $2$ to $3$ unit-cell structures. However, the free-fluid shear stress transfer to pore fluid shear stress is a microscopic phenomenon. The influence of the stress is ranging over
less than one pore-structure; the region over which velocity decays to Darcy's velocity in Fig.~\ref{fig:DNS-results-geomA-compMOD}$b$ extends over distance of $z \approx 0.05\,H\ \refAll{ = 0.5\,l}$ containing only half of one unit-cell structure.
Over this short distance, all the fluid shear stress is transferred over to the solid skeleton, thus contributing to large shear region in Fig.~\ref{fig:DNS-results-solid-displ-lines-depth}$a$.

\begin{figure}
  \begin{center}
  \includegraphics[width=0.6\linewidth]{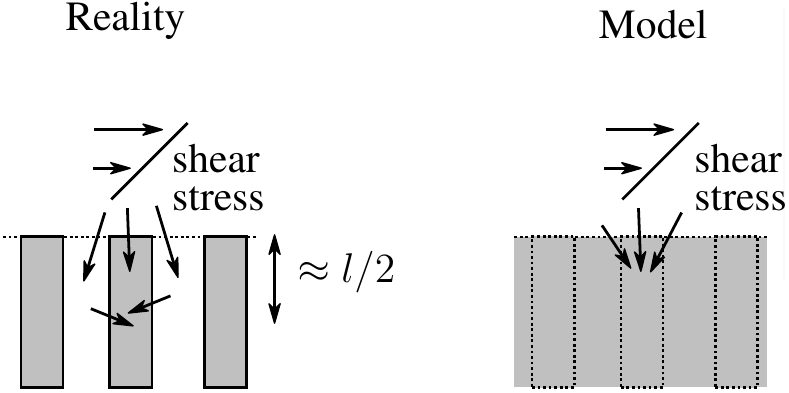} 
  \end{center}
  \caption{\refAll{
  In the left side of the figure we illustrate how the free fluid shear stress
  (sketched with linearly increasing horizontal velocity profile) in reality is
  transferred to both solid skeleton (sketched as grey obstacles) and to pore fluid.
  Due to viscous dissipation, the pore fluid shear stress is then transferred to
  solid skeleton over microscale length $\approx l/2$. On the right side of the figure
  we show that in the model equations all of the free fluid shear stress is transferred
  directly into the solid skeleton at the artificial interface between poroelastic region
  and free fluid region.
\label{fig:sketch-shear-stress-transfer}}}
\end{figure}

As a final comment,
the effective properties ($\ten{C}, \ten{\alpha}$ and $\mathcal{E}$) of the interior can be used at the interface for the stress boundary condition (\ref{eq:homog-v-bc-if}).
The reason is that the shear stress
\refAll{from the free fluid}
is eventually borne only by solid skeleton.
\refAll{As illustrated in Fig.~\ref{fig:sketch-shear-stress-transfer}, left,
from the microscopic -- or
``reality'' -- point of view, the free fluid shear stress is transferred to both solid skeleton
and pore fluid. However, roughly a half of a microstructure below the interface,
the shear stress of the pore fluid
is transferred to the surrounding solid skeleton via viscous friction.}
Therefore the elasticity tensor from the interior is a reasonable estimate for the elasticity tensor at the interface. In other words, the model (Fig.~\ref{fig:sketch-shear-stress-transfer}, right),
in which free fluid shear stress is transferred directly to solid skeleton only,
is a reasonable
approximation.
This is, however, not the case for the velocity boundary
conditions (\ref{eq:dim-vel-if-condition}) and (\ref{eq:dim-vel-if-condition-normal}),
for which the micro-scale viscous dissipation has to be modelled by Navier-slip term ($\ten{L}$);
otherwise, as shown by L\={a}cis \& Bagheri\cite[tab. 2]{lacis2016framework}, the predicted
interface velocity would be around \refAll{$1/\ord$} times smaller compared to DNS.
This is due to the fact that the pore fluid
undergoes a rapid acceleration caused by the free fluid shear at the interface, which renders the interior velocity a very inaccurate estimate for the interface velocity.
%

\subsection{Capturing anisotropic effects with the effective model}
\begin{figure}
  \begin{center}
  \includegraphics[width=1.0\linewidth]{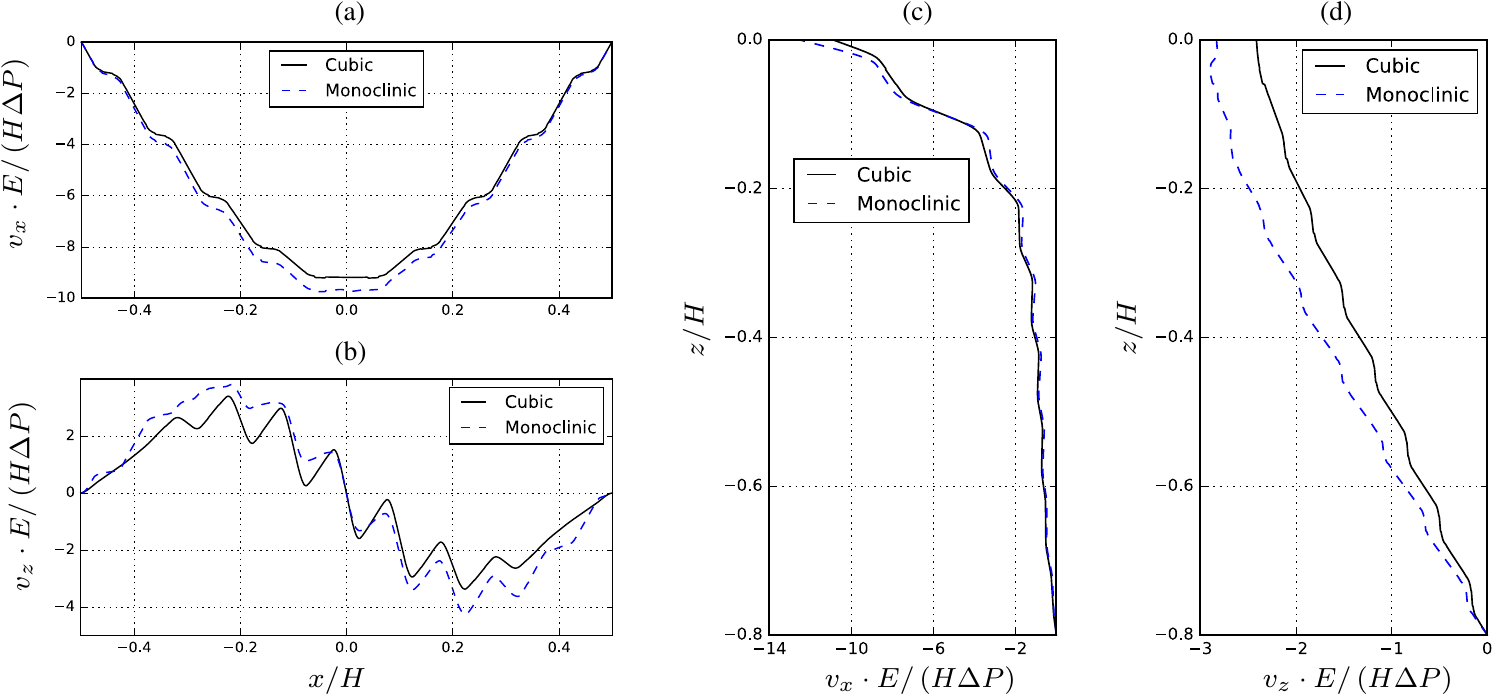} 
  \end{center}
  \caption{
Comparison between DNS
solid displacement data for cavity
with poro-elastic bed
constructed using cubic-symmetric and monoclinic-symmetric geometries.
In the top left frame (a) we show horizontal displacement
near the tip of the solid skeleton at $z =-0.05\,H$. In the bottom left frame
(b) we show the vertical displacement at the same coordinate $z =-0.05\,H$.
In the middle frame (c) we show the horizontal displacement variation
over the vertical coordinate at $x = 0.15\,H$. In the right frame (d) we show
the vertical displacement variation over the vertical coordinate at $x = 0.15\,H$.
\label{fig:geomAB-compDNS-displ}}
\end{figure}

The displacement fields of the two geometries (cubic-symmetric and monoclinic-symmetric) obtained from DNS are compared in Fig.~\ref{fig:geomAB-compDNS-displ}. For the monoclinic-symmetric geometry the horizontal displacement is roughly $6\%$ larger, and the vertical displacement is roughly $10\%$ larger (compared to cubic-symmetric geometry). This difference cannot be explained by the change of porosity only (around $1\%$), therefore the reason must be the introduced anisotropy of the pore geometry.  Moreover, in the horizontal and the vertical displacement fields one can observe a small asymmetry between left and right half of the cavity, which is a sign of symmetry breaking due to anisotropic effects.

In section~\ref{sec:results-cavity}\ref{sec:sub:results-effective}\ref{sec:subsub:results-effective-displ},
a difference was also observed between the two microstructures in the effective model, which was explained by comparing the entries of the effective elasticity tensor.  Fig.~\ref{fig:geomAB-compDNS-displ}, obtained using fully resolved simulations, can be compared with the corresponding figure obtained from the effective model (Fig.~\ref{fig:geomAC-compMOD-displ}). We observe that the relative change between those two geometries is accurately captured by the effective model. This indicates that using the interior effective elasticity, when imposing stress boundary conditions at the interface, provides an effective model that is practically useful in capturing differences in the macroscale, caused by anisotropy within the microstructure.




\section{Discussion} \label{sec:discuss}

%

The numerical framework presented  here is based on first solving a set of microscale problems from section~\ref{sec:effective-prop} to compute the effective tensors ($\ten{C}$, $\ten{K}$, $\ten{\alpha}$, $\ten{K}^\scr{if}$ and $\ten{L}$), followed by solving the macroscopic effective equations coupled through interface conditions to the free fluid (section~\ref{sec:homog-eq}). The equations formed in both these steps stem from multi-scale homogenization approach
and rest on a number of assumptions,
similarly as explained in literature\cite{whitaker1988levels,wood2009role} for the method of volume averaging. 
%
%
These assumptions restrict the range of physical parameters that we are able to model using effective continuum theory. It is therefore essential to understand if the admissible range of parameters are relevant for solving physical problems that arise in nature and engineering. In this section we discuss the necessary assumptions used in the development of the current  framework and how they relate to physical constraints.

The first and foremost prerequisite is scale separation. Based on the numerical tests in the
literature\cite{lacis2016framework} and section~\ref{sec:results-cavity},
we have determined the practical limit on the scale separation parameter
\begin{equation}
\ord = \frac{l}{H} \lesssim 0.1,
\label{eq:ass1}
\end{equation}
which is less restrictive than the asymptotic limit $l / H \rightarrow 0$.
In other words, one does not need to be in the asymptotic limit in order to apply the current framework. 
We compared the effective model results with predictions
from fully resolved simulations in section~\ref{sec:results-cavity} and
observed good agreement with respect to the flow and the displacement fields for
the scale separation parameter $l/H = 0.1$.
It is interesting to point out that Auriault\cite{auriault2005filtration}
has shown the next order corrections to the Darcy's law to be zero
for macroscopically homogeneous porous media,
therefore the Darcy's law holds well also in the case of poor scale separation.
This would explain the
good agreement between the model and DNS for the flow velocity.
It is possible that
similar conclusion could be drawn also from the correctors of the elasticity problem.


We assume that the pore Reynolds number is smaller than one, i.e.
%
\begin{equation}
\frac{\rho_\scr{f} U^d H}{\mu} \leq  \ordest{1}.
\label{eq:ass2}
\end{equation}
Here, recall that the $U^d \equiv l^2 \Delta P / \left(\mu H \right)$ is the
definition of seepage velocity.
The above assumption is a good one for poroelastic materials that are densely packed, resulting in a slow flow through the pores, which can be described using steady linear Stokes equations -- such as those formulated for $\ten{K}$, $\ten{K}^\scr{if}$ and $\ten{L}$ in equations (\ref{eq:unit-perm-prob-1}--\ref{eq:unit-perm-prob-2}), (\ref{eq:intf-unit-perm-prob-1}--\ref{eq:intf-unit-perm-prob-bc-K}) and (\ref{eq:intf-unit-perm-prob-5}--\ref{eq:intf-unit-perm-prob-bc-L}). 
%
%
For larger pore Reynolds numbers the microscopic problems will become non-linear, which renders the current multi-scale approach, based on linear decomposition, unfeasible.
A possible workaround is to use some kind of linearisation, similarly as done
by Zampogna \& Bottaro\cite{bottaro2016rigfibre}.

Next, we make an assumption on how the macroscopic free fluid  time scale $\Delta \tau$ that force the poroelastic bed is related to the microscopic time scale $l/U^d$ inside the bed. Specifically, the frequency $1/\Delta \tau$, at which the free fluid interacts with the poroelastic bed, has to satisfy,
\begin{equation}
\frac{1}{\Delta \tau} \frac{l}{U^d} \leq \ordest{1}. \label{eq:ass3}
\end{equation}
This essentially states that the changes in the free fluid
must be slower compared
to the time a fluid parcel needs to travel a pore length $l$. Thus, from the microscale viewpoint, the external macroscopic forcing is slow, which in turn has the consequence that the corresponding microscale problems (\ref{eq:unit-perm-prob-1}--\ref{eq:unit-perm-prob-2}), (\ref{eq:intf-unit-perm-prob-1}--\ref{eq:intf-unit-perm-prob-bc-K}) and
(\ref{eq:intf-unit-perm-prob-5}--\ref{eq:intf-unit-perm-prob-bc-L})
are steady.
%
If external forcing frequency is higher than $U_d/l$, the Stokes problems (\ref{eq:unit-perm-prob-1}--\ref{eq:unit-perm-prob-2}),
(\ref{eq:intf-unit-perm-prob-1}--\ref{eq:intf-unit-perm-prob-bc-K}) and (\ref{eq:intf-unit-perm-prob-5}--\ref{eq:intf-unit-perm-prob-bc-L}) will become time-dependent, which in turn would require solving convolution integrals in order to take into account the time history, see Mei \& Vernescu\cite[eq. 6.6.11]{mei2010homogenization}.


Another restriction of the current method is on the relative size of the characteristic normal stresses of the flow and the  solid
skeleton. That is, we assume that the macroscale global $\Delta P$ relates to the
characteristic Young modulus of the solid skeleton as follows,
\begin{equation}
\frac{\Delta P}{E} \leq \ordest{\ord}.
\label{eq:ass4}
\end{equation}
This assumption holds in many engineering configurations, where elastic properties of materials are often of order MPa or GPa, while the pressure difference generated by moving fluids is commonly of order
kPa (in wind tunnel experiments, for example). It also holds for many biological systems; 
the elasticity moduli are of order MPa or kPa\cite{carter2003modelling,
kim2012hummingbird,schwager2013functional} where biological materials are often
exposed to much slower fluid flows\cite{gemmell2016dynamic} and consequently
smaller fluid forces.
%

The final requirement is that inertial effects of the solid skeleton at
the microscopic level are small. 
We expect that the
pore-flow is sufficiently slow such that inertial
effects of the solid skeleton in the microscale are not excited.
This implies
a relation between the characteristic
macroscopic pressure, the solid
density and characteristic time scale:
%
\begin{equation}
\frac{\rho_s\, l\, H}{\Delta P \, \Delta \tau^2} \leq \ordest{1}.
\label{eq:ass5}
\end{equation}
%
This results in linear, time-independent microscale solid test problems
(\ref{eq:unit-chi-prob-1}--\ref{eq:unit-chi-prob-2}) and
(\ref{eq:unit-eta-prob-1}--\ref{eq:unit-eta-prob-2}). If this restriction
is not obeyed, the test problems should be complemented using pore-scale
inertial effects. 

Note that the above assumptions are not unique and in order to understand  mathematically why these particular choices are made, we refer the reader to appendix~\ref{app:sec:app-eq-assume}. 
%
At the end, we want to understand whether the resulting effective equations governing the poroelastic bed can be used to describe the length and time scales that are physically relevant. 
In order to do so, we compare restrictions, involving the external macroscopic forcing time scale $\Delta \tau$ (\ref{eq:ass3},\ref{eq:ass5}), with the intrinsic poroelastic time scale $\Delta \tau_\scr{p}$ and the time scale $\Delta \tau_\scr{i}$ related  to the speed of waves in the effective bed. The former time-scale $\Delta \tau_\scr{p}$ characterizes the time for the pressure field to equilibrate via fluid transport in the medium and therefore determines how fast the poroelastic bed can respond to external forcing \cite{skotheim2005physical}.  The latter time scale $\Delta \tau_\scr{i}$ characterizes the time for wave propagation, and therefore also the time it takes for information to propagate through the effective bed.
%
%
Following Skotheim \& Mahadevan\cite{skotheim2005physical}, the  time scales are given by
\[\Delta \tau_\scr{p} \sim \mu H^2 / \left( k C_\scr{eff} \right)\qquad\textrm{and}\qquad \Delta \tau_\scr{i} \sim H
\sqrt{\rho_\scr{s} / C_\scr{eff}},\]
where $k$ is the characteristic permeability and $C_\scr{eff}$ is the characteristic effective elasticity coefficient. 
%
These time scales can be compared to the assumptions discussed above using estimates $k \sim l^2$ and $C_\scr{eff} \sim \theta_s E$, where $\theta_s = 1 - \theta$ is the solid volume fraction. Using assumptions (\ref{eq:ass3}) and (\ref{eq:ass4})
from the current work, we then arrive to
inequality $\Delta \tau \geq \theta_s \Delta \tau_\scr{p}$.
This shows that the macroscopic driving force time scale can be of the same order as the poroelastic material time scale, therefore
the model derived in the current work allows for
description of poroelastic effects in the medium. Furthermore,
using assumptions (\ref{eq:ass4}) and (\ref{eq:ass5}),
we arrive to inequality $\Delta \tau \geq \sqrt{\theta_s} \Delta \tau_\scr{i}$.
This indicates that the model explained in this work allows for description of travelling
waves through poroelastic medium.
To sum up, the current multi-scale approach should allow for the description of problems where $\Delta \tau$ is of the order of $\Delta \tau_p$ (to capture strong fluid-elasticity interaction \cite{skotheim2005physical}) and also of the order of $\Delta \tau_i$
(to capture travelling waves or elastic instabilities).
%

\section{Conclusions} \label{sec:concl}

We have considered the problem of a free-fluid interacting with a poroelastic bed, by deriving and validating an effective continuum model for the bed and its interface with the above free fluid. Although the effective equations of the interior of the bed are well-established, their coupling to a non-trivial vortical free fluid through a set of interface conditions have not been considered and validated before. The imposed interface conditions are (i) the velocity boundary condition for the free fluid; (ii) the pressure continuity boundary condition for the pore-pressure; and (iii) the stress continuity boundary condition for the displacements of poroelastic media. The first two conditions have been derived from first-principles using certain assumptions, while the stress boundary condition is postulated \emph{a posteriori}. In particular, for condition (iii) the interior effective  parameters ($\ten{C}$ and $\ten{\alpha}$) were used, whereas for condition (i) the interface permeability and the Navier-slip tensor were computed. This asymmetry in the treatment of the boundary condition is motivated by the fact that for the velocity boundary condition -- which is needed to solve for the free-fluid -- the transfer of shear-stress to pore flow requires a new Navier-slip tensor $\ten{L}$, because Darcy's law cannot accommodate any shear. Therefore, the interfacial velocity condition requires a special treatment. For the stress boundary condition -- which is needed to solve for the displacement in the bed --
both normal and tangential stresses induced by the free flow above can be matched by the corresponding stresses of the solid.
We have shown that using condition (iii), based on the interior coefficients, provides a satisfactory model; on the one hand, it does result in a -- small, but not insignificant -- discrepancy with the fully-microscopic simulations that we use as a validation; on the other hand, it captures the effects of small changes in the microstructure anisotropy correctly and predicts the overall behaviour in a physically consistent and controllable manner. We thus believe that this approach for modelling the interaction of poroelastic beds with freely moving fluids is a viable framework for engineers. The practical limits of the derived model have been discussed, where we show that the proposed model can be employed for any physically relevant poroelastic material. The corresponding codes of the numerical implementation used in the present work have been released as an open-source
software\cite{github2016UgisShervin}.

In the future, we will further improve the model by treating the stress boundary condition using appropriate interface-cells.
We also want to validate the proposed model for unsteady flows, where inertia is not negligible. Finally, the next step for the model development is to extend it to significant surface deformations, that would require moving the interface
\refAll{and porosity variation in space}.

\enlargethispage{20pt}

\dataccess{The core of the codes used for producing the plots in the current
paper are available as open source software in Github
repository\cite{github2016UgisShervin}. The same repository contains the
flow and displacement fields of DNS simulations presented in this paper.}

\aucontribute{The idea of the present work was conceived through
discussions between
all authors.
U.L. has
derived the homogenized equations with feedback from all other authors.
The numerical implementation
of the homogenized and microscale equations was carried out by U.L.
The reported numerical
simulations were performed by U.L. with feedback from S.B. 
The results from microscale solvers were independently
verified by G.A.Z. using
OpenFOAM for cubic-symmetric material.
Results were analysed and
paper was written by U.L. and S.B. with feedback from G.A.Z.}

\competing{Authors declare no competing interests.}

\funding{The work by U.L and S.B. on this topic has been funded by the
Swedish Research Council (VR-2014-5680) and the G{\"o}ran Gustafsson foundation.}

\ack{G.A.Z. acknowledges several insightful interactions with Prof. A. Bottaro.}

\disclaimer{The views and opinions expressed in this article are those of the authors.}

\bibliographystyle{rspublicnatwithsort}
\bibliography{/home/ugis/Documents/references_UgisL}


%
%
%
%
%

\newpage

\setcounter{page}{1}

{\large Supplementary online appendices for ``A computational continuum model of poroelastic beds''} \\

U. L\={a}cis$^{1}$, G. A. Zampogna$^{2}$ and S. Bagheri$^{1}$

{\flushleft $^{1}$Linn\'{e} Flow Centre, Department of Mechanics KTH, SE-100 44 Stockholm, Sweden\\
$^{2}$DICCA, Scuola Politecnica, Universit\`{a} di Genova, via Montallegro 1,
16145 Genova, Italy}

\appendix

\section{Governing micro-scale equations and assumptions} \label{app:sec:app-eq-assume}

In the current appendix, we explain the non-dimensionalization of
the pore-scale governing
equations, as well as required assumptions and scale estimates for the derivation
of the macroscale governing equations and accompanying pore-scale test
problems. The governing equations, presented in main paper
(\ref{eq:NStokes-ms-dimensional-1}--\ref{eq:solid-fluid-ms-dimensional-bcs}), are
summarized here for convenience:
\begin{align}
\rho_\scr{f} \left( \pdt u_i  + u_j u_{i,j} \right) & = \Sigma_{ij,j}, &  \Sigma_{ij} & = -p \delta_{ij} + 2 \mu \str{}{ij}{u}  & \mbox{in } \Omega_f, \\
u_{i,i} & = 0, & &  & \mbox{in } \Omega_f, \\
u_i & = \pdt v_i, & &  & \mbox{on } \Gamma_s,  \label{app:eq:gov-eq-bc1}\\
\Sigma_{ij} \hn_j & = \sigma_{ij} \hn_j, & &  & \mbox{on } \Gamma_s, \\
\rho_\scr{s} \pdt^2 v_i  & = \sigma_{ij,j}, & \sigma_{ij} & = C^{\textrm{sk}}_{ijkl} \str{}{kl}{v}  & \mbox{in } \Omega_s,
\end{align}
where we have used $\Sigma_{ij}$ and $\sigma_{ij}$ to denote fluid and solid stress
tensors, respectively.
We have also introduced
strain rate tensor for the fluid $\str{}{ij}{u} = 0.5 \left(u_{i,j} + u_{j,i} \right)$ and
strain tensor for the solid $\str{}{ij}{v} = 0.5 \left(v_{i,j} + v_{j,i} \right)$.
Here $\Omega_f$ is the fluid domain (both free fluid and fluid between solid skeleton
-- pore fluid),
$\Omega_s$ is the solid skeleton domain and $\Gamma_s$ is boundary between
solid and fluid. 
For convenience, we
use the index notation, where summation is carried
out over repeating indices and comma
indicates a derivative.
Now we assume that there is scale separation between two
length scales, $l$ and $H$,
with $\ord = l/H \ll 1$.
Other parameters in this problem are $\rho_\scr{f}$, $\rho_\scr{s}$, $\Delta P$,
$\mu$, $\Delta \tau$ and $\ten{C}^\scr{sk}$, where $\Delta P$ is a characteristic pressure difference, $\Delta \tau$
is a characteristic time
scale of the processes being studied and $\ten{C}^\scr{sk}$ is linear solid skeleton
elasticity tensor. Based on these parameters, we can define a velocity scale
$U^d \equiv l^2 \Delta P / \left( \mu H \right)$ arising from momentum balance
at the microscale\cite[eq. A 1]{lacis2016framework}.
We estimate the scales of
flow, pressure, displacement fields and derivatives in the interior as
\begin{equation}
\tilde{u}_i \sim U^d,
\qquad \tilde{p} \sim \Delta P, \qquad \tilde{v}_i \sim l,
\qquad \left( \right)_{,i} \sim \frac{1}{l}, \qquad \pdt \left( \right) \sim \frac{1}{\Delta \tau}.
\label{eq:app:var-order-estim}
\end{equation}
Based on these estimates, we choose to
render the equations dimensionless (we temporarily use  ``tilde'' to denote dimensional
quantities),
using the relationships
\begin{equation}
\tilde{p} = \Delta P\, p,\ \  \ \  \tilde{u}_i = U^d u_i,\ \  \ \ 
\tilde{x}_i = l\, x_i,\ \ \ \ \tilde{t} = \Delta \tau \,t,\ \  \ \ 
\tilde{v}_i =  l\, v_i. \label{eq:app:non-dim-relation}
\end{equation}
Note that non-dimensionalization using physical parameters $\Delta P$, $\mu$, $l$,
$\Delta \tau$ and $H$ is one particular choice and different options are possible.
With this choice we have the dimensionless
order of all terms close to unity
\begin{equation}
u_i = \ordest{1}, \qquad u_{i,j} = \ordest{1}, \qquad p = \ordest{1},
\qquad v_i = \ordest{1}, \qquad \pdt \left( \right) = \ordest{1}.
\label{eq:app:var-order-estim-dimless}
\end{equation}
These orders are required later on, when the multi-scale expansion is
carried out.
Under the chosen normalization (\ref{eq:app:non-dim-relation}), the governing equations
become
\begin{align}
\ord^2 Re_d \left( \bar{\bar{f}}\, \pdt u_i  + u_j u_{i,j} \right) & = \Sigma_{ij,j}, &  \Sigma_{ij} 
& = -p \delta_{ij} + 2 \ord \str{}{ij}{u}  & \mbox{in } \Omega_f, \label{app:eq:fluid-dimless-moment} \\
u_{i,i} & = 0, & &  & \mbox{in } \Omega_f, \label{app:eq:fluid-dimless-cont} \\
u_i & = \bar{\bar{f}}\, \pdt v_i, & &  & \mbox{on } \Gamma_{s},  \label{app:eq:nond-gov-eq-bc1}\\
\Sigma_{ij} \hn_j & = \frac{1}{\ord} \sigma_{ij} \hn_j, & &  & \mbox{on } \Gamma_{s}, \label{app:eq:nond-gov-eq-bc2} \\
\ord \bar{\bar{\rho}} \pdt^2 v_i  & = \frac{1}{\ord} \sigma_{ij,j},
& \sigma_{ij} & = \bar{\bar{E}} C^{\textrm{sk}}_{ijkl} \str{}{kl}{v}  & \mbox{in } \Omega_s. \label{app:eq:solid-dimless-displ}
\end{align}
This system of non-dimensional equations is valid everywhere and so
far there are no additional assumptions employed.
The total list of physical parameters in the problem is $\rho_\scr{f}$,
$\rho_\scr{s}$, $\Delta P$, $\mu$, $l$, $H$,
$\Delta \tau$, and $\ten{C}^\scr{sk}$. This adds up to $7 + n$ parameters,
where $n$ depends on the properties of elastic skeleton, and for the current
work $n = 2$ (we assume that it is isotropic and elasticity is characterized
using $2$ parameters). According to Buckingham $\pi$ theorem, we then have to
have 6 dimensionless parameters, which are
\begin{equation}
Re_d = \frac{\rho_f U^d H}{\mu},\ \ \bar{\bar{f}} = \frac{1}{\Delta \tau} \frac{l}{U^d}, \ \ \ord = \frac{l}{L},\ \ \bar{\bar{E}} = \frac{\ord E}{\Delta P}, \ \  C^{\textrm{sk}}_{ijkl} = \frac{\tilde{C}^{\textrm{sk}}_{ijkl}}{E}, \ \ \bar{\bar{\rho}} = \frac{\rho_s\, l\, H}{\Delta P \, \Delta \tau^2}.
\label{eq:app:dimless-numbers}
\end{equation}
Here, we have separated a Young modulus out of the skeleton elasticity and the
dimensionless tensor $C^\scr{sk}_{ijkl}$ is characterized by only one scalar
-- Poisson's ratio $\nu$.
If the skeleton is built from an anisotropic material, then the
dimensionless tensor $C^\scr{sk}_{ijkl}$ would imply a larger number of non-dimensional
parameters. For the isotropic case, this form is useful to
describe behaviour of isotropic materials with similar 
Poisson's ratio but different
Young's modulus.
In addition, since the resulting equation system in the Stokes limit is
linear, in the main paper we plot displacements normalized with dimensionless
elasticity parameter $v_i \cdot \bar{\bar{E}}$, which in dimensional setting
becomes $\tilde{v}_i \cdot E / \left( H \Delta P \right)$. In other words, the
values presented in the displacement plots would correspond to simulation
with $\bar{\bar{E}} = 1$.

To be able to use the multi-scale expansion, one has to estimate the relative orders
of all the terms.
We assume the following:
\begin{align}
\bar{\bar{E}} = \ordest{ 1 }, \qquad \bar{\bar{\rho}} = \ordest{ 1 }, \qquad Re_d \leq \ordest{ 1 },
\qquad \bar{\bar{f}} = \ordest{ 1 }. \label{app:eq:eq-material-assume}
\end{align}
Now the relative magnitude between different terms is completely illustrated
by the $\ord$ pre-factors in the governing non-dimensional equations
(\ref{app:eq:fluid-dimless-moment}--\ref{app:eq:solid-dimless-displ}) in
the interior of the poroelastic material. 
This, however, does not hold in the free fluid.
In other words, in the free fluid region, it is not possible to
use the Darcy's law as a governing equation.

In the main paper section~\ref{sec:discuss} introduced restrictions of this theory
has been obtained by relaxing the equality conditions
from (\ref{app:eq:eq-material-assume}). By ``relaxing'' we mean
that the theory should not only be applicable to one value of each of the
dimensionless parameter, as set by (\ref{app:eq:eq-material-assume}),
but for a range of dimensionless parameters. The generalization from the
equality assumptions (\ref{app:eq:eq-material-assume}) to the inequality restrictions
is done as fallows:
\begin{enumerate}
  \item The equality assumption
$\bar{\bar{E}} = \ordest{ 1 }$
is relaxed to the inequality restriction based on
numerical tests in the main paper section~\ref{sec:results-cavity},
in which one could see that
the theory works also for very stiff materials $\bar{\bar{E}} \gg \ordest{ 1 }$.
Therefore we conclude that the theory works for a range of dimensionless
parameters $\bar{\bar{E}} \geq \ordest{ 1 }$, which includes the theoretical
assumption $\bar{\bar{E}} = \ordest{ 1 }$ and numerical validation
at $\bar{\bar{E}} \gg \ordest{ 1 }$.
  \item The equality assumption
$\bar{\bar{\rho}} = \ordest{ 1 }$ is relaxed to inequality restriction
$\bar{\bar{\rho}} \leq \ordest{ 1 }$ based on physical understanding that
the developed method works also in the non-inertial regime; that is, when the solid
density is sufficiently small that inertial effects can be neglected, the
parameter $\bar{\bar{\rho}}$ can be set to zero.
  \item Finally, the equality assumption
$\bar{\bar{f}} = \ordest{ 1 }$ is relaxed to inequality restriction
$\bar{\bar{f}} \leq \ordest{ 1 }$ based on
numerical tests in the main paper section~\ref{sec:results-cavity}.
There we have validated the model equations for an ``infinitely slow''
or steady test problem, for which $\bar{\bar{f}} = 0$. Uniting the two
parameter limits, for which the theory should work,
we get $\bar{\bar{f}} \leq \ordest{ 1 }$.
\end{enumerate}


Although the equality assumptions on the dimensionless parameters
(\ref{app:eq:eq-material-assume}) are strict and in principle one could
set them to the specific value in the equations, we retain these
coefficients in the equations to facilitate tests away from the outlined
assumptions.


\section{Derivation of homogenized effective equations} \label{app:sec-app:deriv}

\renewcommand{\theequation}{B \arabic{equation}}

We are now ready to derive the homogenized governing equations for the
poroelastic material in macroscale. Additionally, we obtain the microscale
test problems for determining the effective material properties.
We introduce the macroscale and microscale coordinates 
\[
X_i=\frac{\tilde{x}_i}{H} \qquad \textrm{and} \qquad x_i=\frac{\tilde{x}_i}{l},
\] 
respectively. These coordinates are appropriate to describe the macroscopic and microscopic variations and are related to each other by $X_i=\epsilon x_i$.
In the new coordinates, 
the spatial derivative is given by
\begin{equation}
\left( \right)_{,i} = \left( \right)_{,i_1} + \ord \left( \right)_{,i_0}, \label{app:eq:mse-chain-der}
\end{equation}
where $\left( \right)_{,i_0}$ denotes the derivative with respect to  $X_i$ and $\left( \right)_{,i_1}$ with respect to  $x_i$. Then we introduce a multi-scale expansions as
\begin{align}
&u_i(X_i,x_i) = u_i^{(0)}(X_i,x_i) + \ord u_i^{(1)} (X_i,x_i) + \ord^2 u_i^{(2)} (X_i,x_i) + \mathcal{O}(\ord^3),\label{app:eq:inter-expand-u}\\
&p(X_i,x_i) =  p^{(0)} (X_i,x_i) + \ord p^{(1)} (X_i,x_i) + \ord^2 p^{(2)} (X_i,x_i) + \mathcal{O}(\ord^3),\label{app:eq:inter-expand-p} \\
&v_i(X_i,x_i) =  v^{(0)}_i (X_i,x_i) + \ord v^{(1)}_i (X_i,x_i) + \ord^2 v^{(2)}_i (X_i,x_i) + \mathcal{O}(\ord^3),\label{app:eq:inter-expand-p}
\end{align}
which we insert into the main
equations (\ref{app:eq:fluid-dimless-moment}--\ref{app:eq:solid-dimless-displ}).
The first two orders of fluid momentum
equation (\ref{app:eq:fluid-dimless-moment}) after the expansion
with corresponding stress tensors are
\begin{align}
\ord^0: & \ \ \vare{0}{\Sigma_{ij,j_1}} = 0, &  & \vare{0}{\Sigma_{ij}} = - \vare{0}{p} \delta_{ij}, \\
\ord^1: & \ \ \vare{1}{\Sigma_{ij,j_1}} + \vare{0}{\Sigma_{ij,j_0}} = 0, & & \vare{1}{\Sigma_{ij}} = - \vare{1}{p} \delta_{ij} + 2 \str{1}{ij}{ \vare{0}{u} }, \label{app:eq:exp-fluid-moment-1}
\end{align}
where we have introduced the strain rate tensor in the micro-scale
$\str{1}{ij}{u} = 0.5 \left(u_{i,j_1} + u_{j,i_1} \right)$. The first two orders
of the fluid continuity
equation (\ref{app:eq:fluid-dimless-cont}) after the expansion are
\begin{align}
\ord^0: & \ \ \vare{0}{u_{i,i_1}} = 0, \label{app:eq:exp-fluid-cont-0} \\
\ord^1: & \ \ \vare{1}{u_{i,i_1}} + \vare{0}{u_{i,i_0}} = 0. \label{app:eq:exp-fluid-cont-1}
\end{align}
The first three orders of the solid
displacement equation (\ref{app:eq:solid-dimless-displ})
after the expansion
with corresponding stress tensors are
\begin{align}
\ord^{-1}: & \ \ \bar{\bar{E}} \vare{-1}{\sigma_{ij,j_1}} = 0, & & \vare{-1}{\sigma_{ij}} =  C^{\textrm{sk}}_{ijkl} \str{1}{kl}{ \vare{0}{v} }, \\
\ord^0: & \ \ \bar{\bar{E}} \vare{0}{\sigma_{ij,j_1}} + \bar{\bar{E}} \vare{-1}{\sigma_{ij,j_0}} = 0, & & \vare{0}{\sigma_{ij}} =  C^{\textrm{sk}}_{ijkl} \str{1}{kl}{ \vare{1}{v} } + C^{\textrm{sk}}_{ijkl} \str{0}{kl}{ \vare{0}{v} },\label{app:eq:exp-solid-1} \\
\ord^1: & \ \ \bar{\bar{E}} \vare{1}{\sigma_{ij,j_1}} + \bar{\bar{E}} \vare{0}{\sigma_{ij,j_0}} = \bar{\bar{\rho}} \vare{0}{\ddot{v}_i}, & & \vare{1}{\sigma_{ij}} =  C^{\textrm{sk}}_{ijkl} \str{1}{kl}{ \vare{2}{v} } + C^{\textrm{sk}}_{ijkl} \str{0}{kl}{ \vare{1}{v} }. \label{app:eq:exp-solid-2}
\end{align}
Here we have used double-dot to denote second derivative in time
$\ddot{v}_i = \pdt^2 v_i$.
Note that the orders of the solid equations are shifted to one order lower value
to match the orders of the stress boundary condition between solid structure and
pore geometry (\ref{app:eq:nond-gov-eq-bc2}). The boundary conditions
(\ref{app:eq:nond-gov-eq-bc1}--\ref{app:eq:nond-gov-eq-bc2}) for all orders
$n \in \left(0, \infty \right)$ take the form
\begin{align}
\vare{n}{u_i} & = \bar{\bar{f}}\, \vare{n}{\dot{v}_i}, \label{app:eq:exp-bc1} \\
\vare{n}{\Sigma_{ij}} \hn_j & =  \bar{\bar{E}} \vare{n}{\sigma_{ij}} \hn_j,
\end{align}
where the time derivative is denoted using dot notation $\dot{v}_i = \pdt v_i$ and there are no lower order values from the fluid stresses $\vare{-1}{\sigma_{ij}} \hn_j = \vare{-1}{\Sigma_{ij}} \hn_j = 0$.

Solving $\ordest{1}$-problem for fluid and $\ordest{\ord^{-1}}$-problem for elasticity, gives us
\begin{equation}
p^{(0)} = p^{(0)} \left( X_i \right), \qquad v^{(0)}_i = v^{(0)}_i\left( X_i \right),
\end{equation}
that is, the leading order pressure and displacement depend
only on the macroscale. The fluid $\ordest{\ord}$-problem is
\begin{align}
-p^{(1)}_{,i_1} + u^{(0)}_{i,j_1j_1} & = p^{(0)}_{,i_0}, \\
u^{(0)}_{i,i_1} & = 0, \\
\left. u^{(0)}_{i} \right|_{\Gamma} & = \bar{\bar{f}} \, \dot{v}^{(0)}_i.
\end{align}
This problem can be solved using the ansatz
\begin{align}
u^{(0)}_{i} & = - \mathcal{K}_{ij} \vare{0}{p_{,i_0}} + \bar{\bar{f}} \, \dot{v}^{(0)}_i, \label{app:eq:lin-assumpt-fluid1} \\
p^{(1)} & = - \mathcal{A}_j \vare{0}{p_{,j_0}}, \label{app:eq:lin-assumpt-fluid2}
\end{align}
where the second term in the velocity ansatz is added to satisfy the velocity
boundary condition at the surface with the solid skeleton. In principle,
one could introduce a second proportionality tensor, but the resulting
problem would always be trivial to solve, yielding factor $\delta_{ij}$
before the velocity of the solid skeleton.
Exactly the same conclusion holds also for the derivation of the
velocity boundary condition \citep{lacis2016framework}, therefore
the time derivative of the displacement appears in the boundary condition
in the main paper (\ref{eq:dim-vel-if-condition}) and
(\ref{eq:dim-vel-if-condition-normal}).
The solid $\ordest{1}$-problem is
\begin{align}
\left[ C^{\textrm{sk}}_{ijkl} \str{1}{kl}{ v^{(1)} } \right]_{,j_1} & = 0, \\
\left. \bar{\bar{E}} \left[ C^{\textrm{sk}}_{ijkl} \str{1}{kl}{ v^{(1)} } + C^{\textrm{sk}}_{ijkl} \str{0}{kl}{ v^{(0)} } \right] \hn_j \right|_{\Gamma} & = - \left. p^{(0)} \hn_i \right|_{\Gamma}.
\end{align}
This problem can be solved using the ansatz
\begin{equation}
v^{(1)}_i  = \chi_{ikl} \str{0}{kl}{ v^{(0)} } - \frac{\eta_{i}}{\bar{\bar{E}}}  p^{(0)}, \label{app:eq:lin-assumpt-solid} \\
\end{equation}
where division by $\bar{\bar{E}}$ is done in order to arrive with test problem
for $\eta_{i}$, which is independent from the dimensionless 
parameter $\bar{\bar{E}}$.

Inserting these ansatzes into the governing equations, one can
group different terms together and form a particular solution by setting individual
group contributions to zero. For the fluid problem, only one Stokes system
has to be solved as a closure problem, which is
\begin{align}
\mathcal{A}_{j,i_1} - \mathcal{K}_{ij,k_1k_1} & = \delta_{ij}, \\
\mathcal{K}_{ij,i_1} & = 0, \\
\left. \mathcal{K}_{ij} \right|_{\Gamma} & = 0, \label{app:eq:mcrosc-interior-permeab-prob-3}
\end{align}
where $\Gamma$ is the boundary with solid skeleton. This problem is solved in one
unit-cell using periodic boundary conditions at all sides for both $K_{ij}$ and $A_{j}$ fields.
This test problem is explained in the main paper near equations
(\ref{eq:unit-perm-prob-1}--\ref{eq:unit-perm-prob-2}) in
dimensional setting.
For the elasticity, two quantities have to be solved for. The first one is for
the displacement tensor
$\chi_{ijk}$, which is governed by
\begin{align}
\left[ C^{\textrm{sk}}_{ijkl} \left\{ \str{1}{kl}{ \chi^{mn} } + \delta_{km} \delta_{ln} \right\} \right]_{,j_1} & = 0, \\
\left[ C^{\textrm{sk}}_{ijkl} \left\{ \str{1}{kl}{ \chi^{mn} } + \delta_{km} \delta_{ln} \right\} \right] \hn_j & = 0,
\end{align}
where elasticity $\bar{\bar{E}}$ has been cancelled out as a common factor. In this case,
the displacement field is also exposed to periodic boundary conditions at all unit-cell sides.
In order to ensure uniqueness of the solution, one can constrain the average
values of the displacement tensors or enforce point constraints. Finally, there is a problem
for proportionality vector before pressure
\begin{align}
\left[ C^{\textrm{sk}}_{ijkl} \str{1}{kl}{ \eta } - \delta_{ij} \right]_{,j_1} & = 0, \\
\left[ C^{\textrm{sk}}_{ijkl} \str{1}{kl}{ \eta } - \delta_{ij} \right] \hn_j & = 0,
\end{align}
where $\bar{\bar{E}}$ has been cancelled in the product of the elasticity tensor
with the pre-factor $\eta / \bar{\bar{E}}$ in the displacement ansatz. Also this problem
needs to be fixed using periodic boundary
conditions and constraints on average values. The elasticity
test problems are explained in the main paper near equations
(\ref{eq:unit-chi-prob-1}--\ref{eq:unit-chi-prob-1}) and
(\ref{eq:unit-eta-prob-1}--\ref{eq:unit-eta-prob-2})
in dimensional setting.

To derive the governing equations,
the volume average operator, as defined in the main paper,
equation (\ref{eq:def-vol-avg}), is used.
We employ the volume averaging on the ansatz of the fluid velocity and arrive to
\begin{equation}
\langle \vare{0}{u_i} \rangle - \theta \bar{\bar{f}} \vare{0}{\dot{v}_i} = - \langle K_{ij}
\rangle \vare{0}{p_{,i_0}}, \label{app:eq:der-relDarcy-law}
\end{equation}
where $\theta = V_f / l^3$ is the volume fraction of the fluid (porosity). Note
that $\vare{0}{\dot{v}_i}$ is independent of the microscale, therefore
averaging only results in $\theta$ pre-factor. In order to continue derivations,
we express the volume average of the first order velocity as
\begin{equation}
\langle \vare{1}{u_{i,i_1}} \rangle = \frac{1}{l^3} \oint \vare{1}{u_i} \hn_i \,\mathit{dS} = \frac{1}{l^3} \int_{\Gamma} \vare{1}{u_i} \hn_i \,\mathit{dS} + \frac{1}{l^3} \int_{\Pi} \vare{1}{u_i} \hn_i \,\mathit{dS}, \label{app:eq:der-macro-mass-1}
\end{equation}
where the closed surface integral has been divided into a part over the 
boundary between solid and fluid $\Gamma$ and a part over the periodic
boundary $\Pi$. Due to periodicity, the second term is zero, and the
first one can be rewritten using boundary condition (\ref{app:eq:gov-eq-bc1}) at the
solid skeleton as
\begin{align}
\frac{1}{l^3} \int_{\Gamma} \vare{1}{u_i} \hn^f_i \,\mathit{dS} & = \frac{\bar{\bar{f}}}{l^3} \int_{\Gamma} \vare{1}{\dot{v}_i} \hn^f_i \,\mathit{dS} = - \frac{\bar{\bar{f}}}{l^3} \int_{\Gamma} \vare{1}{\dot{v}_i} \hn^s_i \,\mathit{dS} - \frac{\bar{\bar{f}}}{l^3} \int_{\Pi} \vare{1}{\dot{v}_i} \hn^s_i \,\mathit{dS} = \nonumber \\
 & = - \frac{\bar{\bar{f}}}{l^3} \oint \vare{1}{\dot{v}_i} \hn^s_i \,\mathit{dS} = - \bar{\bar{f}} \langle \vare{1}{\dot{v}_{i,i_1}} \rangle,  \label{app:eq:der-macro-mass-2}
\end{align}
where we have added an integral of zero value, taken over the
periodic boundaries of the solid.
We have also transferred between adjacent domains with $\hn^s_i = -\hn^f_i$. Hence in this
example we clearly see that the application of Gauss theorem in the
microstructure relies on
the assumption of periodicity. Nevertheless, the resulting model could also be used if
the periodicity assumption
is satisfied only approximately.
Using the linear expression for the first order displacement (\ref{app:eq:lin-assumpt-solid}), we arrive with the continuity equations for the macroscale
\begin{equation}
\langle \vare{0}{u_i} \rangle_{,i_0} = \bar{\bar{f}} \langle \chi_{i,i_1}^{kl} \rangle \str{0}{kl}{
\vare{0}{\dot{v}} } - \frac{\bar{\bar{f}}}{\bar{\bar{E}}} \langle \eta_{i,i_1} \rangle \vare{0}{\dot{p}},
\label{app:eq:der-conserv-mass}
\end{equation}
where we have used the assumption that the porosity is uniform, and thus the
derivative and volume averaging commute. To obtain the final equation in
the macroscale, we introduce a stress tensor, which is defined
over the whole volume as
\begin{equation}
\vare{0}{T_{ij}} = \left\{ \begin{array}{lr}
\vare{0}{\Sigma_{ij}} & \mbox{in } \Omega_f, \\
\bar{\bar{E}} \vare{0}{\sigma_{ij}} & \mbox{in } \Omega_s,
\end{array} \right.
\end{equation}
which can be differentiated with respect to the macroscale, and the result is
\begin{equation}
\vare{0}{T_{ij,j_0}} = \left\{ \begin{array}{lr}
\vare{0}{\Sigma_{ij,j_0}} & \mbox{in } \Omega_f, \\
\bar{\bar{E}} \vare{0}{\sigma_{ij,j_0}} & \mbox{in } \Omega_s.
\end{array} \right.
\end{equation}
Now we average the divergence of the total stress tensor
\begin{equation}
\langle \vare{0}{T_{ij,j_0}} \rangle = \langle \vare{0}{\Sigma_{ij,j_0}} \rangle + \langle \bar{\bar{E}} \vare{0}{\sigma_{ij,j_0}} \rangle =  - \langle \vare{1}{\Sigma_{ij,j_1}} \rangle + \langle \bar{\bar{\rho}} \vare{0}{\ddot{v}_i} \rangle -  \langle \bar{\bar{E}} \vare{1}{\sigma_{ij,j_1}} \rangle,
\end{equation}
where we have used equations (\ref{app:eq:exp-fluid-moment-1}) and (\ref{app:eq:exp-solid-2}) to rewrite the divergence of zeroth order stress tensors. In the same way as in equations (\ref{app:eq:der-macro-mass-1} -- \ref{app:eq:der-macro-mass-2}), we can use Gauss theorem, periodicity and the boundary conditions, to write:
\begin{equation}
\langle \vare{1}{\Sigma_{ij,j_1}} \rangle = - \langle \bar{\bar{E}} \vare{1}{\sigma_{ij,j_1}} \rangle.
\end{equation}
On the other hand, the zeroth order stress tensors can be averaged using explicit expressions, for the fluid stress we get
\begin{equation}
\langle  \vare{0}{\Sigma_{ij,j_0}} \rangle = - \theta \vare{0}{p_{,j_0}} \delta_{ij},
\end{equation}
whereas for the solid stress we find
\begin{equation}
\langle \bar{\bar{E}} \vare{0}{\sigma_{ij,j_0}} \rangle = \langle \bar{\bar{E}} C^{\textrm{sk}}_{ijkl} \left[ \str{1}{kl}{\chi^{mn}} + \delta_{km}\delta_{ln} \right] \rangle \left[ \str{0}{mn}{ \vare{0}{v} } \right]_{,j_0} - \langle C^{\textrm{sk}}_{ijkl} \str{1}{kl}{\eta} \rangle {p_{,j_0}}.
\end{equation}
To sum up, the final equation in the macroscale is
\begin{equation}
\left(1 - \theta \right) \bar{\bar{\rho}}\, \vare{0}{\ddot{v}_i} = \left[ \bar{\bar{E}} C_{ijmn} \str{0}{mn}{ \vare{0}{v} } \right]_{,j_0} - \alpha_{ij} \vare{0}{p_{,j_0}},
\end{equation}
where the effective elasticity tensor is
\begin{equation}
C_{ijmn} = \langle C^{\textrm{sk}}_{ijkl} \str{1}{kl}{\chi^{mn}} \rangle + \left( 1 - \theta \right) C^{\textrm{sk}}_{ijmn}, \label{app:eq:effective-elast-def}
\end{equation}
and the tensor in front of the pore pressure is
\begin{equation}
\alpha_{ij} = \theta \delta_{ij} + \langle C^{\textrm{sk}}_{ijkl} \str{1}{kl}{\eta} \rangle.
\end{equation}

The equations to be solved for the homogenized macroscale model expressed
using relative Darcy's law (\ref{app:eq:der-relDarcy-law}) in
the conservation of mass law (\ref{app:eq:der-conserv-mass}) are
\begin{align}
\left[ \frac{\mathcal{E}}{E} \right] \vare{0}{\dot{p}} - \left[ \left( \mathcal{K}_{ij} \frac{l^2}{\mu} \right) \vare{0}{p_{,j}} \right]_{,i} & = - \alpha_{kl} \str{}{kl}{ \vare{0}{\dot{v}} }, \\
\left(1 - \theta \right) \rho_s \vare{0}{\ddot{v}_i} & = \left[ \left( E C_{ijmn} \right) \str{}{mn}{ \vare{0}{v} } \right]_{,j} - \alpha_{ij} \vare{0}{p_{,j}},
\end{align}
where all variables and coordinates are now dimensional. These are the
equations explained in the main paper (\ref{eq:dim-pore-momentum})
and (\ref{eq:dim-pore-pressure}). We note that if the solid
density is very small, as allowed by the restriction
(\ref{eq:ass5}) in the main paper, then
the inertial term in solid displacement field can
be neglected. Here we have also used the equality
\begin{equation}
\alpha_{kl} = \theta \delta_{kl} - \langle \chi_{i,i_1}^{kl} \rangle, \label{app:eq:ten-alph-def2}
\end{equation}
which has been derived by
Mei \& Vernescu\cite[p. 258--259]{mei2010homogenization} and confirmed
within the current work numerically. Now we can relate the parameters given by dimensionless
unit-cell problems to the parameters used in main paper (denoted using tilde) as
\begin{equation}
\tilde{\mathcal{E}} = \frac{\mathcal{E}}{E}, \qquad \tilde{K}_{ij} = K_{ij}\,l^2
\qquad \mbox{and} \qquad \tilde{C}_{ijkl} = E C_{ijkl}.
\end{equation}
The pore pressure contribution tensor $\alpha_{ij}$ is dimensionless
in both settings. Note
that one can formulate the unit-cell problems also in
dimensional setting, as written in the
main paper.
The microscale problems in the associated open-source software
repository\cite{github2016UgisShervin}
are dimensionless, and one should use the relationships above in order to
obtain coefficients for the dimensional equations.

\section{Obtaining all model parameters using elasticity tensors} \label{app:sec-app:subderiv-from-elast}

\renewcommand{\theequation}{C \arabic{equation}}

In the current section, we demonstrate how to arrive to all necessary elastic properties
using only elasticity of solid skeleton and effective elasticity of solid skeleton,
similarly as employed by Gopinath \& Mahadevan\cite{gopinath2011elastohydrodynamics}.
First, we define an inverse of fourth-rank tensor $\ten{A}$ in index notation as
\begin{equation}
A^{-1}_{ijkl} A_{klmn} = \delta_{im} \delta_{jn}.
\end{equation}
In order to obtain pore pressure contribution tensors using elasticity matrices, we start
by multiplying effective elasticity tensor (\ref{app:eq:effective-elast-def})
by the inverse
of the skeleton elasticity and obtain
\begin{equation}
C^{-1(\scr{sk})}_{ijkl} C_{klmn} = \left\langle \frac{1}{2} \left(\chi^{mn}_{i,j_1} 
+ \chi^{mn}_{j,i_1} \right)
\right\rangle + (1-\theta) \delta_{im} \delta_{jn},
\end{equation}
which we then multiply by the identity matrix from left to get
\begin{equation}
\delta_{ij} C^{-1(\scr{sk})}_{ijkl} C_{klmn} = \left\langle \chi^{mn}_{i,i_1}
\right\rangle + (1-\theta) \delta_{mn}.
\end{equation}
Now, in order to match the expression for the tensor $\alpha_{ij}$ (\ref{app:eq:ten-alph-def2}),
we subtract the obtained result from the identity matrix
\begin{equation}
\delta_{mn} - \delta_{ij} C^{-1(\scr{sk})}_{ijkl} C_{klmn} = \theta \delta_{mn}
- \left\langle \chi^{mn}_{i,i_1}
\right\rangle,
\end{equation}
which due to symmetry in elasticity tensors $C_{ijkl} = C_{klij}$ can be rewritten
as
\begin{equation}
\delta_{mn} - C_{mnkl} C^{-1(\scr{sk})}_{klij} \delta_{ij} = \theta \delta_{mn}
- \left\langle \chi^{mn}_{i,i_1}
\right\rangle,
\end{equation}
which is the final expression for the pore-pressure contribution tensor. In tensorial
notation, this expression can be written using double contraction operator as
\begin{equation}
\ten{\alpha} = \ten{\delta} - \ten{C} : \ten{C}^{-1}_\scr{sk} : \ten{\delta}.
\label{eq:app:alpha-vs-elast}
\end{equation}
This expression is exactly the same one as reported by Gopinath \&
Mahadevan\cite[eq. 2.4]{gopinath2011elastohydrodynamics}.
Carrying out similar derivation for the coefficient $\mathcal{E}$, one can obtain
\begin{equation}
\mathcal{E} = \ten{\delta} : \left( \ten{C}^{-1}_\scr{sk} : \left[ \ten{\alpha} - \theta \ten{\delta} \right] \right).
\label{eq:app:e-vs-elast}
\end{equation}
Therefore if one is successful in finding the
effective elasticity tensor by other means,
the pore-pressure contribution tensor $\alpha$ and coefficient $\mathcal{E}$
can be recovered using expressions (\ref{eq:app:alpha-vs-elast}--\ref{eq:app:e-vs-elast}).

\end{document}